\documentclass{article} 
\usepackage[twocolumn]{emulateapj}
\usepackage{epsf,psfig}

\def\gmc{g cm$^{-3}$}
\def\kms{km s$^{-1}$}
\def\rhoc{$\rho_{\rm c}$}
\def\ms{M$_\odot$}
\def\mr{$M_r$}
\def\ni{$^{56}$Ni}
  
\def\mdot{$\dot M$}

\def\e#1{$\times$ $10^{#1}$ }
\def\ee#1{$10^{#1}$ }

\def\mm#1{$M_{\rm #1}$}

\def\ltsima{$\; \buildrel < \over \sim \;$}
\def\ltsim{\lower.5ex\hbox{\ltsima}}
\def\gtsima{$\; \buildrel > \over \sim \;$}
\def\gtsim{\lower.5ex\hbox{\gtsima}}
\def\etal{et al. }
\def\afoe{$\times$ 10$^{51}$ ergs }

\begin{document}

\submitted{Published in the Astrophysical Journal Supplement Series,
1999, v.125, pp.439-462}

\title{NUCLEOSYNTHESIS IN CHANDRASEKHAR MASS MODELS FOR TYPE IA
SUPERNOVAE AND CONSTRAINTS ON PROGENITOR SYSTEMS AND BURNING FRONT PROPAGATION}

\author{\footnote[1]{Present Address: 
Department of Physics, Nihon University, Tokyo 101-8308, Japan
}Koichi Iwamoto$^{1,4,5}$, Franziska Brachwitz$^2$, 
Ken'ichi Nomoto$^{1,4,5}$, Nobuhiro Kishimoto$^1$, \\
Hideyuki Umeda$^{4,5}$, 
W. Raphael Hix$^{3,5}$, Friedrich-K. Thielemann$^{2,3,5}$}
\affil{$^1$ Department of Astronomy, University of Tokyo, Tokyo 113-0033, Japan}
\affil{$^2$ Department of Physics and Astronomy, Universitty of Basel,
CH-4056 Basel, Switzerland}
\affil{$^3$ Oak Ridge National Laboratory, Oak Ridge, TN 37831-4576, USA}
\affil{$^4$ Research Center for the Early Universe, School of 
Science, University of Tokyo, Tokyo 113-0033, Japan}
\affil{$^5$ Institute for Theoretical Physics, Santa Barbara, CA 93106-4030,
USA}

\begin{abstract}

	The major uncertainties involved in the Chandrasekhar mass
models for Type Ia supernovae (SNe Ia) are related to the companion
star of their accreting white dwarf progenitor (which determines
the accretion rate and consequently the carbon ignition density) and the 
flame speed after the carbon ignition.  We calculate explosive
nucleosynthesis in relatively slow deflagrations with a variety of
deflagration speeds and ignition densities to put new constraints on
the above key quantities.  The abundance of the Fe-group, in
particular of neutron-rich species like $^{48}$Ca, $^{50}$Ti,
$^{54}$Cr, $^{54,58}$Fe, and $^{58}$Ni, is highly sensitive to the
electron captures taking place in the central layers. The
yields obtained from such a slow central deflagration, and from a fast
deflagration or delayed detonation in the outer layers, are combined
and put to comparison with solar isotopic abundances.  To avoid
excessively large ratios of $^{54}$Cr/$^{56}$Fe and
$^{50}$Ti/$^{56}$Fe, the central density of the "average" white dwarf
progenitor at ignition should be as low as \ltsim 2 \e9 \gmc.  To
avoid the overproduction of $^{58}$Ni and $^{54}$Fe, either the flame
speed should not exceed a few \% of the sound speed in the central low
$Y_e$ layers, or the metallicity of the average progenitors has to be
lower than solar.  Such low central densities can be realized by a
rapid accretion as fast as $\dot M$ \gtsim 1 $\times$
10$^{-7}$M$_\odot$ yr$^{-1}$.  In order to reproduce the solar
abundance of $^{48}$Ca, one also needs progenitor systems that undergo
ignition at higher densities.  Even the smallest laminar flame speeds
after the low-density ignitions would not produce sufficient amount of
this isotope.  We also found that the total amount of $^{56}$Ni, the
Si-Ca/Fe ratio, and the abundance of some elements like Mn and Cr
(originating from incomplete Si-burning), depend on the density of the
deflagration-detonation transition in delayed detonations.  Our
nucleosynthesis results favor transition densities slightly below
2.2$\times 10^7$~g cm$^{-3}$.

\end{abstract}

\section {Introduction}

	There are strong observational and theoretical indications
that Type Ia supernovae (SNe Ia) are thermonuclear explosions of
accreting white dwarfs (e.g., Wheeler et al. 1995; Nomoto, Iwamoto \&
Kishimoto 1997; Branch 1998).  Theoretically, both the Chandrasekhar
mass white dwarf models and sub-Chandrasekhar mass models have been
considered (see, e.g., Arnett 1996; Nomoto \etal 1994, 1996a, 1997b,
1997c; Canal, Ruiz-Lapuente, \& Isern 1997 for reviews of recent
progress).  Though these white dwarf models can account for the basic
observational features of SNe Ia, the exact binary evolution that
leads to SNe Ia has not been identified yet.  Various evolutionary
scenarios have been proposed, which include (1) a double degenerate
scenario, i.e., the merging of two C+O white dwarfs in a binary system
with a combined mass exceeding the Chandrasekhar mass limit (e.g.,
Iben \& Tutukov 1984; Webbink 1984) and (2) a single degenerate
scenario, i.e., accretion of hydrogen or helium via mass transfer from
a binary companion at a relatively high rate (e.g., Nomoto 1982a).  In
the case of helium accretion at low rates, He detonates at the base of
the accreted layer before the system reaches the Chandrasekhar mass
(Nomoto 1982b; Woosley \& Weaver 1986, 1994a; Livne \& Arnett 1995).
Currently, the issues of the Chandrasekhar mass versus
sub-Chandrasekhar mass models and the double degenerate versus single
degenerate scenarios are still debated (see, e.g., Renzini 1996 and
Branch et al. 1995 for recent reviews), but they are being confronted
with an increasing number of observational constraints.

	The observational search for the double degenerate scenario
led to the discovery of a few binary white dwarfs systems, but with
combined mass being smaller than the Chandrasekhar mass (Renzini
1996).  Theoretically, it has been suggested that the merging of
double white dwarf systems leads to accretion-induced collapse rather
than to SNe Ia (Nomoto \& Iben 1985; Saio \& Nomoto 1985, 1998).  The
Chandrasekhar versus sub-Chandrasekhar mass issue has recently
experienced some progress. Photometric and spectroscopic features of
SNe Ia in early phases clearly indicate that Chandrasekhar mass models
give a much more consistent picture than the sub-Chandrasekhar mass
models of helium detonations (e.g., H\"oflich \& Khokhlov 1996; Nugent
\etal 1997). This leaves us with the most likely progenitor system, a
single degenerate system with hydrogen accretion from the companion
star, leading to a Chandrasekhar-mass white dwarf. However, the
Chandrasekhar mass model W7 (Nomoto, Thielemann, \& Yokoi 1984;
Thielemann, Nomoto, \& Yokoi 1986), widely used in galactic chemical
evolution calculations, may require improvements in terms of the
Fe-group composition because it predicts significantly higher
$^{58}$Ni/$^{56}$Fe ratios than solar.  The direct determination of Ni
abundances in late time SN Ia spectra is therefore important
(Ruiz-Lapuente 1997; Liu, Jeffery, \& Schultz 1997; Mazzali et
al. 1998).  The recent findings of supersoft X-ray sources, being
potential progenitors of SN Ia events with high accretion rates,
causing ignition at low densities (van den Heuvel \etal 1992;
Rappaport, Di Stefano, \& Smith 1994; Di Stefano \etal 1997), leave
hope for Chandrasekhar-mass models, which meet all these requirements,
also for late time spectra.

	The presupernova evolution of an accreting white dwarf depends
on the accretion rate $\dot M$, the composition of the material
transferred from the companion star, and the initial mass of the white
dwarf (e.g., Nomoto 1982a; Nomoto \& Kondo 1991).  Chandrasekhar mass
white dwarfs can be obtained with a relatively high mass transfer rate
of hydrogen of the order $\dot M$ $\approx$ 4 \e{-8} -- \ee{-5}
M$_\odot$ yr$^{-1}$. At $\dot M > 4 \times 10^{-6}$ M$_\odot$
yr$^{-1}$, the accreting white dwarf blows off a strong wind, which
reduces $\dot M$ to an effective accretion rate below \ee{-6}
$M_\odot$ yr$^{-1}$ (Hachisu, Kato, \& Nomoto 1996, 1999a).  This
avoids the formation of an extended envelope in the accreting white
dwarf.  (Nomoto , Nariai, \& Sugimoto 1979).  At such rates hydrogen
and helium burn steadily or with weak flashes, leading to a white
dwarf with a growing C+O mass.

	For the Chandrasekhar mass white dwarf model, carbon ignition
in the central region leads to a thermonuclear runway. The ignition
density depends on the stellar structure as a function of previous
accretion history.  High accretion rates lead to higher central
temperatures, i.e.  favoring lower ignition densities.  A flame front
then propagates at a subsonic speed $v_{\rm def}$ as a {\sl
deflagration wave} owing to heat transport across the front (Nomoto et
al. 1984). The major and yet not fully solved questions are related to
the propagation of the burning front. Timmes \& Woosley (1992) have
analyzed the propagation speed of laminar burning fronts in one
dimension as a function of density and fuel composition.  However, the
propagation in three dimensions is influenced by instabilities that
can enhance the effective radial flame speed beyond its laminar value.

	The flame front is subject to various types of instabilities,
namely, thermal instabilities (Bychkov \& Liberman 1995a), the
Landau-Darrius (L-D) instability (Landau \& Lifshitz 1987), the
Rayleigh-Taylor (R-T) instability, and the Kevin-Helmholtz (K-H)
instability (Niemeyer, Woosley, \& Hillebrandt 1996). The turbulent
burning regime associated with the R-T bubbles on global scales has
been studied (Livne 1993; Arnett \& Livne 1994a; Khokhlov 1995;
Niemeyer \& Hillebrandt 1995a; Niemeyer et al. 1996), but there remain
many uncertainties, related partially to numerical resolution but also
to the role and spectrum of turbulent length scales (Hillebrandt \&
Niemeyer 1997).  With the present uncertainties, it is essential to
perform parameterized sets of calculations that explore the possible
range of effective radial flame speeds.

	In the deflagration wave, electron captures enhance the
neutron excess. The amount of electron capture depends on both $v_{\rm
def}$ (influencing the time duration of matter at high temperatures,
and with it the availability of free protons for electron capture and
the high-energy tail of the electron energy distribution) and the
central density of the white dwarf $\rho_9 = \rho_{\rm c}$/10$^9$ g
cm$^{-3}$ (increasing the electron chemical potential).  The resultant
nucleosynthesis in slow deflagrations (see, e.g., Khokhlov 1991b) has
some distinct features compared with faster deflagrations like W7
(Nomoto \etal 1984; Thielemann \etal 1986), thus providing important
constraints on these two parameters.  The constraint on the central
density is equivalent to a constraint on the accretion rate, as
discussed above.  After an initial deflagration in the central layers,
the deflagration is accelerated as in W7, or assumed to turn into a
detonation at lower densities, as in the delayed detonation models
(Khokhlov 1991a; Woosley \& Weaver 1994a). For the latter, the
uncertain transition density $\rho_{\rm tr}$ would result in a variety
of total masses of $^{56}$Ni and expansion velocities of the outer
layers.

	To obtain constraints on the three parameters ($\rho_{\rm
ign}$, $v_{\rm def}$, and $\rho_{\rm tr}$), we performed explosive
nucleosynthesis calculations for slow deflagrations followed by a
delayed detonation or a fast deflagration. These calculations assumed
spherical symmetry and therefore might not be fully adequate for a
realistic and consistent approach, but we expect that they lead at
least to some clues how abundance features relate to the spherical
average of these quantities in realistic models.  Initially slower
deflagrations cause an earlier expansion of the outer layers with
respect to the arrival of the burning front (as information of the
central ignition propagates with sound speed; Nomoto, Sugimoto, \& Neo
1976) and lead to low densities for the outer deflagration and
detonation layers.  In such a case, even a detonation does not lead to
a pure Fe-group composition, as expected in central detonations, and
intermediate mass elements from Si to Ca are produced at a wide range
of expansion velocities.  Therefore, if the deflagration-detonation
transition (DDT) density is well tuned, delayed detonations can meet
these observational requirements as well as fast deflagration
(Kirshner \etal 1993).  Compared with the earlier delayed detonation
models by Khokhlov (1991b), we adopt a larger and more detailed
nuclear reaction network that alos includes electron screening.  In
comparison to the fast deflagration models by Woosley (1997b), we use
initial models with lower central densities and smaller flame speeds,
to concentrate on our main aim, which is to find the "average" SN Ia
conditions responsible for their nucleosynthesis contribution to
galactic evolution, i.e., especially the Fe-group composition.

	From the very early days of explosive nucleosynthesis
calculations, when no direct connection to astrophysical sites was
possible yet, it was noticed (Trimble 1975) that the solar Fe-group
composition could be reproduced with a superposition of matter from
explosive Si burning with about 90\% originating from a $Y_e$ = 0.499
source and 10\% from a $Y_e$ = 0.46 source.  As it has been shown that
SN II ejecta with $Y_e <$ 0.498 could cause serious problems in
comparison with observations (Thielemann, Nomoto, \& Hashimoto 1996),
SNe Ia have to be identified with this second source and the
appropriate conditions which lead to the best agreement with solar
abundances. For this reason we present detailed yields of delayed
detonation models as well as fast deflagration models and compare them
with solar abundances for a number of "training sets" of ignition
densities, flame speeds, and DDT densities.

	These nucleosynthesis constraints can provide clues to the
explosion mechanism (i.e., the speed of the burning front) and the
ignition density (i.e., the accretion rate from the binary companion)
for the "average" or dominating SN Ia contributions during galactic
chemical evolution. This is the purpose of the present paper.  We have
to be aware that there are some systematic variations that manifest
themselves in light curves (i.e., the brighter one is slower; Phillips
et al. 1990; Hamuy \etal 1995) and might lead to a variation in
nucleosynthesis as well (H\"oflich \& Khokhlov 1996, H\"oflich,
Wheeler \& Thielemann 1998). It is further of importance to explore
metallicity effects, which might have an influence on the evolution of
the progenitor systems, with respect to the initial mass function
(IMF) and composition of white dwarfs as well as the binary accretion
history (e.g., Yoshii, Tsujimoto, \& Nomoto 1996; H\"oflich et
al. 1998; Umeda et al. 1999; H\"oflich et al. 1999). Only the latter
will clarify whether the nature of SNe Ia at high redshifts is the
same as for nearby SNe Ia, which enters the determination of the
cosmological parameters $H_0$ and $q_0$ (e.g., Branch \& Tammann 1992;
Riess, Press, \& Kirshner 1995; Riess et al. 1999; Perlmutter \etal
1997, 1999). Observable spectral features could possibly help to
identify the metallicity due to slight changes in nucleosynthesis
(H\"oflich et al. 1998; Hatano et al. 1999; Lentz et al. 1999).

	After a description of our model calculations in \S2,
including initial models, the hydrodynamic treatment, and a discussion
of the input physics, we present in \S3 detailed nucleosynthesis
results from slow deflagrations and delayed detonations in comparison
with the carbon deflagration model W7.  In \S4, the integrated
abundances of SNe Ia models are combined with those of SNe II to
compare with solar abundances.  Finally, we discuss constraints on
possible evolutionary scenarios and give conclusions in \S5.  Very
preliminary accounts of the present investigations on nucleosynthesis
in slow deflagrations and delayed detonations have been given in
Nomoto et al. (1997c) and Thielemann et al. (1997).

\section {Initial Models, Explosion Hydrodynamics, and Input Physics}

\subsection{Initial Models}

	We adopt two models with central densities of $\rho_9$ = 1.37
(C) and 2.12 (W) at the onset of thermonuclear runaway, i.e., at the
stage when the timescale of the temperature rise in the center becomes
shorter than the dynamical timescale.  Here C and W imply that these
are the same models as calculated for C6 and W7, respectively (Nomoto
\etal 1984).  The initial white dwarfs of these models, before the
onset of H-accretion, have a mass of $M$ = 1.0 \ms, a central
temperature of $T_c$ = 1.0 \e7 K, and compositions of $X(^{12}$C) =
0.475, $X(^{16}$O) = 0.50, $X(^{22}$Ne) = 0.025.

	The outer layers of the mass grid extend to the steady
hydrogen burning shell as an outer boundary. The temperature and
density at the burning shell are determined from the boundary
condition that the accreted matter is processed into helium with the
mass accretion rate \mdot.  These values increase from initally 8 \e7
K and 1 \e4 \gmc~ to 1 \e8 K and 1 \e6 \gmc~ at the point of central
carbon ignition. The accretion rate for case C is due to steady and
stable hydrogen burning corresponding to the C+O core increase during
an asymptotic giant branch evolution (see Nomoto 1982a). The rate for
case W is kept constant up to the point of carbon ignition at the
center. The exact values are given by equations (1) and (2), where $M$
indicates the mass of the accreting white dwarf.

\begin{eqnarray}
  {\dot M}({\rm C})& = &8.5 \times 10^{-7} (M/{\rm M}_\odot - 0.52)
             {\rm M}_\odot {\rm y}^{-1}\\
  {\dot M}({\rm W})& = & 4 \times 10^{-8} {\rm M}_\odot {\rm y}^{-1}.
\end{eqnarray}

	During the accretion phase the white dwarf mass $M$ increases
with time and the central temperature increases as a result of heat
inflow from the H-burning shell as well as compressional heating.
Cooling is due mostly to plasmon neutrino losses and neutrino
bremsstrahlung.  Cooling due to Urca shells and the convective Urca
process is not taken into account.  This has no effect on case C, but
could delay the ignition in case W to higher densities (e.g.,
Paczynski 1973; Iben 1982; Nomoto \& Iben 1985; Barkat \& Wheeler
1990).  When the central density \rhoc~ reaches 1.5 \e9 \gmc~ (C) or
2.5 \e9 \gmc~ (W), carbon is ignited in the center, where the nuclear
energy generation rate exceeds the neutrino losses.  When the central
temperature increases owing to carbon ignition, a convective core
develops. The convective energy transport is calculated in the
framework of the time-dependent mixing length theory (Unno 1968). At
$T_{\rm c} \sim$ 8 \e8 K convection can no longer transport energy in
our model and the central region undergoes a thermonuclear runaway
with \rhoc~ = 1.37 \e9 \gmc~ (C) or 2.12 \e9 \gmc~ (W).

\subsection {Slow Deflagrations}

\begin{figure}[ht]
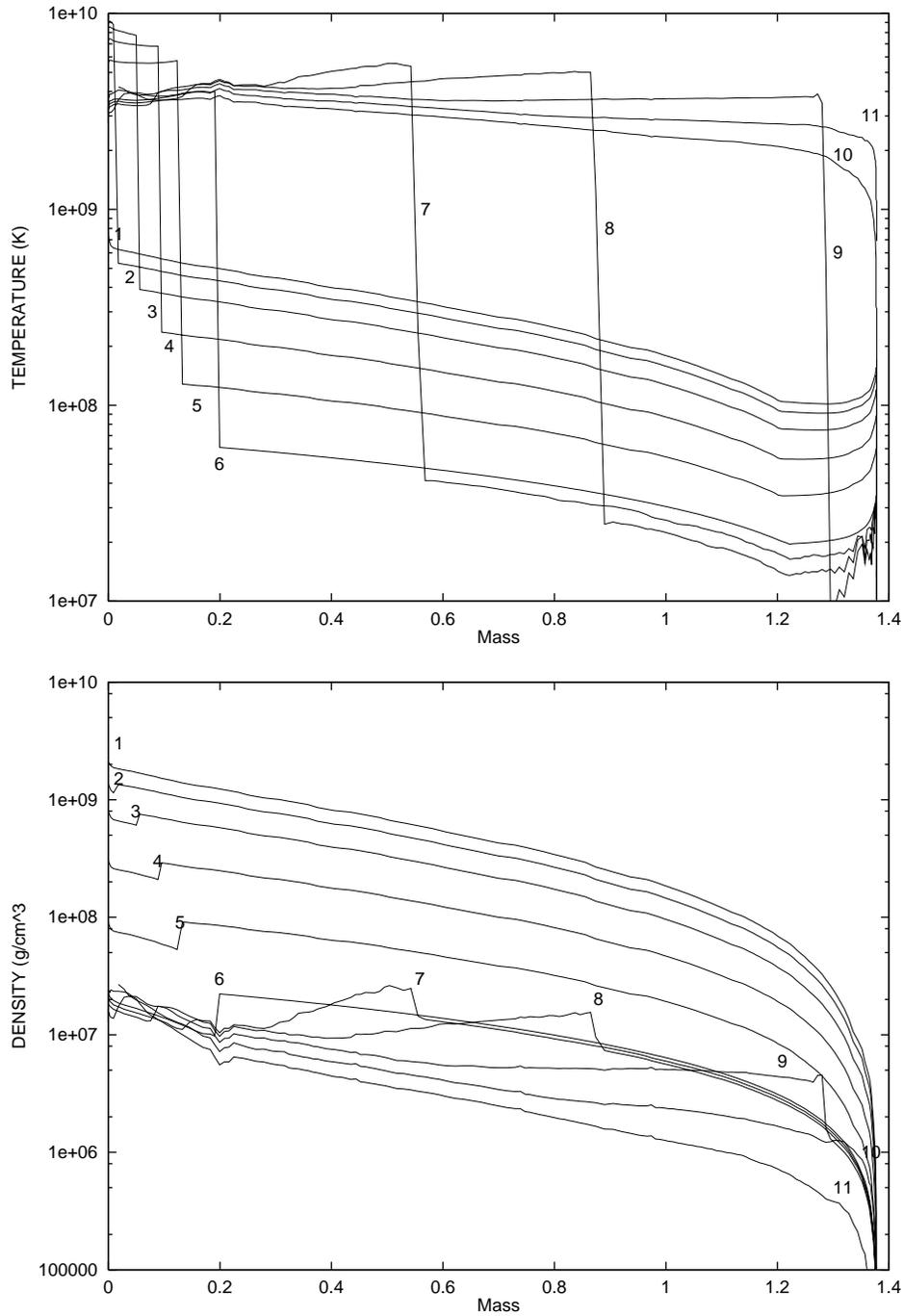

\epsfxsize=7cm
\epsfysize=6cm
\vspace*{-9.5cm}
\epsfbox[-47 61 351 755]{tmr2.epsi}
\epsfxsize=7cm
\epsfysize=6cm
\vspace{-3cm}
\epsfbox[-47 61 351 755]{rmr2.epsi}
\vspace{10cm}
\caption{
Temperatures and densities for the deflagration wave (WS15) and
delayed detonation (DD1) as a function of radial mass coordinate
\mr. The labels indicate different instances in time, 1-6 correspond
to the deflagration phase, 7-11 to the propagation of the detonation after
the deflagration-detonation transition. \label{defr}}
\end{figure}

	We know the absolute lower limit for the deflagration speed
after central ignition from the one-dimensional analysis of laminar
flame fronts (Timmes \& Woosley 1992), being close to 1\% of the local
sound speed $v_s$. On the other hand, any hydrodynamic instability can
enhance the speed. Our parametrized "fast" deflagration studies, which
reached 10\%-30\% of the sound speed and produced the model W7 (Nomoto
\etal 1984; Thielemann \etal 1986), however, resulted in problematic
Fe-group nucleosynthesis (see also discussion below). The flame speed
found in multidimensional hydrodynamic simulations is still subject
to large uncertainties, ranging from $v_{\rm def}/v_{\rm s} \sim$
0.015 (Niemeyer \& Hillebrandt 1995a) to $v_{\rm def}/v_{\rm s} \sim$
0.1 (Niemeyer et al. 1996).  Therefore, it is important to investigate
how the nucleosynthesis outcome depends on the flame speed. In order
to contrast our "fast" deflagration model W7, we study "slow"
deflagrations here and choose the following parameter ranges: After
the central thermonuclear runaway, we assume that a slow (S)
deflagration propagates with speeds $v_{\rm def}/v_{\rm s}$ = 0.015
(WS15, CS15), 0.03 (WS30, CS30), and 0.05 (CS50) and consider also
extreme cases of fully and initially laminar flame fronts (WLAM, WSL). 
The location of the deflagration wave in radial mass coordinate \mr\
and the changes in temperature and density are shown in Figure
\ref{defr} as a function of time.  Behind the deflagration wave, the
temperature rises quickly to values as high as $T$ = 9 $\times$ \ee9 K
and the material experiences nuclear statistical equilibrium (NSE).
As the flame front propagates outward, the white dwarf expands slowly
which reduces the central density \rhoc.  The decrease in \rhoc~ for
these slow deflagrations is significantly slower than in W7.


\placefigure{defr}

\subsection {Transition from Deflagration to Detonation}

	If the deflagration speed continues to be much slower than in
W7, the white dwarf undergoes a large amplitude pulsation, as first
found by Nomoto \etal (1976).  In this {\sl pulsating deflagration}
model, the white dwarf expands and nuclear burning is quenched when
the total energy of the star is still negative.  In the following
contraction more material burns, resulting in a positive total energy
$E$.  Eventually the white dwarf is completely disrupted.  The model
by Nomoto \etal (1976) resulted in $E$ = 5 \e{49} ergs and a \ni~ mass
of \mm{Ni}~ $\sim$ 0.15 M$_\odot$.  Such a pulsating deflagration
produces explosion energies too small to account for typical SNe Ia
but might be responsible for rare events such as SN 1991bg.

	In order to produce sufficient amounts of radioactive
$^{56}$Ni ($\sim$ 0.6 $M_\odot$) to power SNe Ia light curves by a
deflagration wave, the flame speed must be accelerated.  The degree to
which the flame speed is increased depends on the effect of R-T
instabilities during the pulsation (Woosley 1997a).  The deflagration
might induce a detonation when reaching the low-density layers.  In
the {\sl delayed detonation} model (Khokhlov 1991a; Woosley \& Weaver
1994b), the deflagration wave is assumed to be transformed into a
detonation at a specific density during the first expansion phase.  In
the {\sl pulsating} delayed detonation model (Khokhlov 1991b), the
transition into a detonation is assumed to occur close to the maximum
compression after recontraction, as a result of mixing.

	Physical mechanisms by which such deflagration-to-detonation
transitions (DDTs) occur have been studied by Arnett \& Livne (1994b),
Niemeyer \& Woosley (1997), Khokhlov, Oran, \& Wheeler (1997), and
Niemeyer \& Kerstein (1997): (1) When a sufficiently shallow
temperature gradient is formed in the fuel, a deflagration propagates
as a result of successive spontaneous ignitions.  Such an over-driven
deflagration propagates supersonically.  (2) If a sufficiently large
amount of fuel has such a shallow temperature gradient, the
deflagration may induce a detonation wave.  The critical masses for
the formation of a detonation are quite sensitive to the carbon mass
fraction $X({\rm C})$, e.g., $\sim$ \ee{-19} \ms~ and $\sim$ \ee{-14}
\ms~ at $\rho \sim$ $3\times$ \ee7 \gmc~ for $X({\rm C})$=1.0 or 0.5,
respectively.  (Niemeyer \& Woosley 1997).  (3) Such a shallow
temperature gradient region in the fuel may be formed if the fuel is
efficiently heated by turbulent mixing with ashes.  Such a mixing and
heat exchange may occur when the turbulent velocity associated with
the flame destroys the flame (Niemeyer \& Woosley 1997; Khokhlov et
al. 1997). Note that whether the DDT occurs by this mechanism is
controversial (Niemeyer 1999), and thus the exact density of the DDT
is still debated (Niemeyer \& Kerstein 1997). Therefore,
nucleosynthesis constraints on the DDT density are important to
obtain.

\begin{figure}[ht]
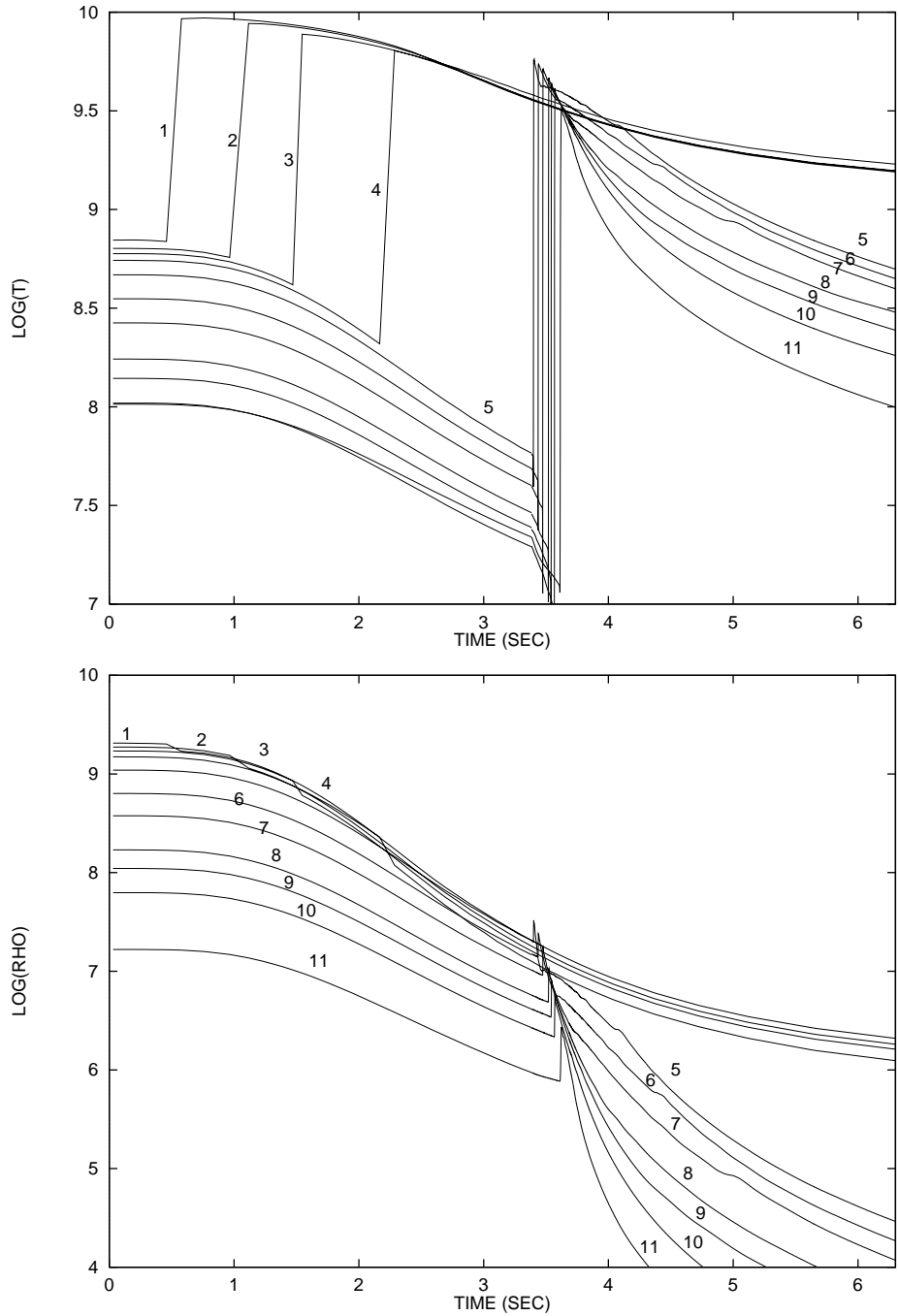

\epsfxsize=7cm
\epsfysize=6cm
\vspace*{-9.5cm}
\epsfbox[-47 61 351 747]{tt2.epsi}
\epsfxsize=7cm
\epsfysize=6cm
\vspace{-3cm}
\epsfbox[-47 61 351 747]{rt2.epsi}
\vspace{10cm}
\caption{
The distribution of temperatures and densities as a
function of time during the passage of the slow deflagration (WS15)
and the subsequent delayed detonation (DD1). The labeled numbers
indicate different mass zones.  The prominent spike, occurring for
labels $\ge$ 5, shows the action of the detonation and its fast
propagation (small time differences between different mass zones).
\label{dens}}
\end{figure}

\begin{figure}[ht]
\epsfxsize=7cm
\epsfysize=6cm
\epsfbox[-890 300 -530 926]{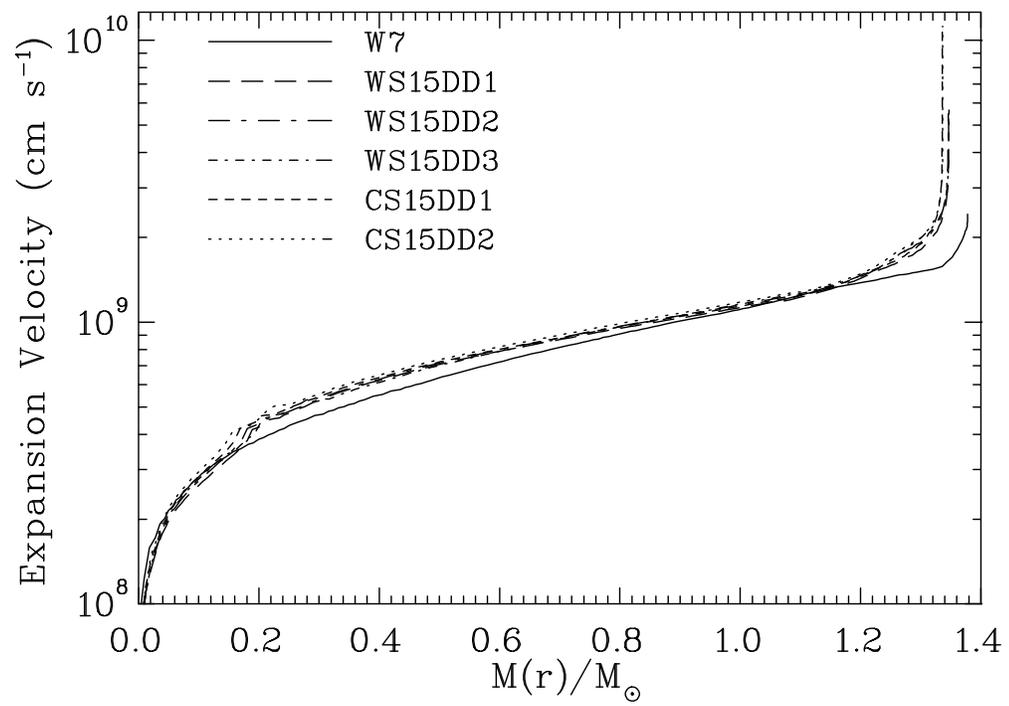}
\vspace{8cm}
\caption{
The distribution of expansion velocities after the passage
of the slow deflagration and the subsequent delayed detonation. 
\label{vexp}}
\end{figure}

	Motivated by the discussion above we transform the slow
deflagrations WS15 and CS15 artificially into detonations when the
density ahead of the flame decreases to 3.0, 2.2, and 1.7 \e7 \gmc~
(DD3, DD2, and DD1, respectively, where 3, 2, and 1 indicate
$\rho_7$=$\rho/10^7$~g~cm$^{-1}$ at the DDT).  Then the carbon
detonation propagates through the layers with $\rho < $ \ee8 \gmc.
Figures \ref{dens} and \ref{vexp} show the density distribution and
expansion velocity after the passage of the slow deflagration and the
subsequent delayed detonation.  The explosion energy $E$ and the mass
of synthesized \ni~ of these WSDD/CSDD models as well as W7 and W70
are summarized in Table 1.  Here W70 is the same hydrodynamical model
as W7 except for the initial mass fractions of $X(^{22}$Ne) = 0.0,
$X(^{12}$C) = 0.50, $X(^{16}$O) = 0.50. This corresponds to zero
initial metallicity because $^{22}$Ne originates from the initial CNO
elements.



	The increase of the $^{56}$Ni mass from W7 to W70 is due to
the fact that the composition in W70 corresponds to a $Y_e$=0.5 or a
proton/neutron ratio of 1, i.e., symmetric matter. In mass zones that
are not affected by electron capture but which undergo complete
Si burning, $^{56}$Ni is then the dominant nucleus, without competition
by more neutron-rich species.  Detonations in lower density matter do
not necessarily lead to complete Si burning. Therefore, the amount of
$^{56}$Ni is smallest in DD1 models and largest in DD3 models. The
small difference between WS and CS DD-models is again a $Y_e$
effect. CS models have a smaller central density, which leads in the
central regions to a smaller number of electron captures and larger
$Y_e$ values, resulting again in more symmetric matter.

\subsection{Input Physics}

	The methods for performing these calculations were the same as
used in Shigeyama \etal (1992) and Yamaoka \etal (1992).  We apply an
implicit Lagrangian hydrodynamics code (Nomoto \etal 1984) for the
slow deflagration and a Lagrangian PPM hydro code, as used in
Shigeyama \etal (1992), for the detonation phase. Both codes use the
same mass grid of 200 radial zones for the white dwarf. The nuclear
reaction network and the reaction rate library utilized are the same
as described in Thielemann et al. (1996), i.e., thermonuclear rates
using the Hauser-Feshbach formalism (Thielemann, Arnould, \& Truran
1987), experimental charged particle rates from Caughlan \& Fowler
(1988), neutron induced rates from Bao \& K\"appeler (1987), and
extensions towards the proton and neutron drip lines from van Wormer
\etal (1994) and Rauscher \etal (1994) with Coulomb enhancement
factors from Ichimaru (1993). Electron capture rates were adopted from
Fuller, Fowler, \& Newman (1980, 1982, 1985). For nuclei beyond
$A$=60, only ground state decay properties were used. We consulted the
analysis of Aufderheide \etal (1994) so that the influence of a
nucleus with a significant impact on $Y_e$ (either via decay or
electron capture) was neglected in either of the conditions
experienced in our calculations. It might, however, require a further
study to test against the sensitivity of these electron capture rates.
The present set of Fuller, Fowler \& Newman (1982) is based on
estimates for average properties of the Gamow-Teller giant resonance
rather than on more secure shell model calculations for fp-shell
nuclei in the Fe-group (Dean et al. 1998).

	One of the problems of nucleosynthesis calculations which
follow a thermonuclear evolution on long timescales through
high-temperature regimes is the lack of accuracy. This lack of
accuracy is due to the fact that in such situations, where actually a
nuclear statistical equilibrium should exist, the cancellation of huge
opposing rate flows is only attained up to machine accuracy (i.e.,
$\approx 10^{-12}-10^{-13}\times$ the term size). In a similar way the
term $1/\Delta t$, appearing in the diagonal of the Jacobi matrix of
the multidimensional Newton-Raphson iteration (Hix \& Thielemann
1999b), can become numerically negligible in comparison to reaction
rate terms in the same sum, which leads to numerically singular
matrices.  For that reason, we have adopted here the accurate
solution, i.e., we followed the nuclear evolution with a screened NSE
network containing 299 species during periods when temperatures beyond
$T$=$6\times 10^9$~K are attained (see Hix \& Thielemann 1996, 1999a).
This takes into account the changes in binding energy or reaction
Q-values due to screening. Weak reaction rates (electron captures and
beta-decays), which are not in an equilibrium and occur on longer
timescales, were included by using the NSE abundances.  In this way we
could track accurately the evolution of the electron fraction $Y_e$.

\section {Explosive Nucleosynthesis}

\subsection {Slow Deflagration}

\begin{figure}[ht]
\epsfxsize=7cm
\epsfysize=6cm
\epsfbox[-42 665 195 749]{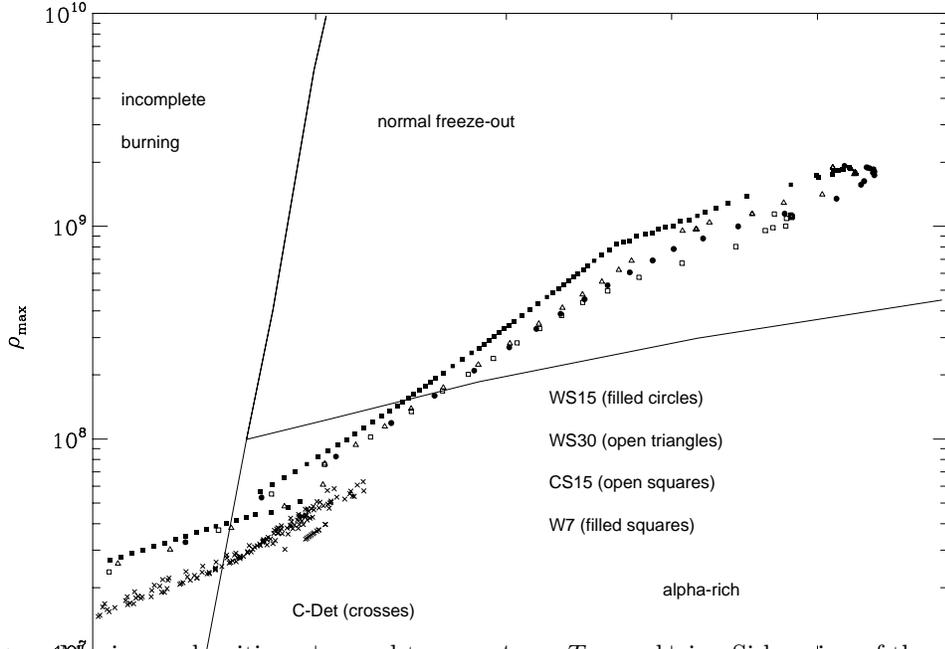}
\vspace{6cm}
\caption{
Maximum densities $\rho_{\rm max}$ and temperatures
$T_{\rm 9,max}$ during Si-burning of the central layers obtained during
the propagation of the burning front.  The model W7 is compared to
three slow deflagrations which start (for a density of 2.1$\times
10^9$g cm$^{-3}$) with burning front velocities of 1.5\% (slow
deflagration WS15) and 3\% (WS30) of the sound speed and for an
ignition density of 1.37$\times 10^9$g cm$^{-3}$ with a burning front
velocity of 1.5\% of sound (CS15). The crosses (C-det) correspond to
carbon detonations of sub-Chandrasekhar mass models. They will not be
discussed here, but it is obvious that the smaller temperatures and
densities lead to negligible amounts of electron
captures. \label{rhoT}}
\end{figure}

	In the present study we follow the approach discussed above by
varying the ignition density and the initial deflagration velocity and
test specifically the effect on the Fe-group composition in the
central part. Figure \ref{rhoT} shows the maximum densities and
temperatures for the inner mass zones experiencing complete and
incomplete Si burning in the models WS15, WS30, CS15, and W7 as a
comparison. It can be recognized that WS15 and WS30 experience
densities similar to that of W7, which is due to the common ignition
densities of initial model W, with WS15 attaining the highest
temperatures. CS15, based on initial model C, is shifted to smaller
densities.

\begin{figure}[hb]
\epsfxsize=7.5cm
\epsfysize=6cm
\epsfbox[-900 100 -450 403]{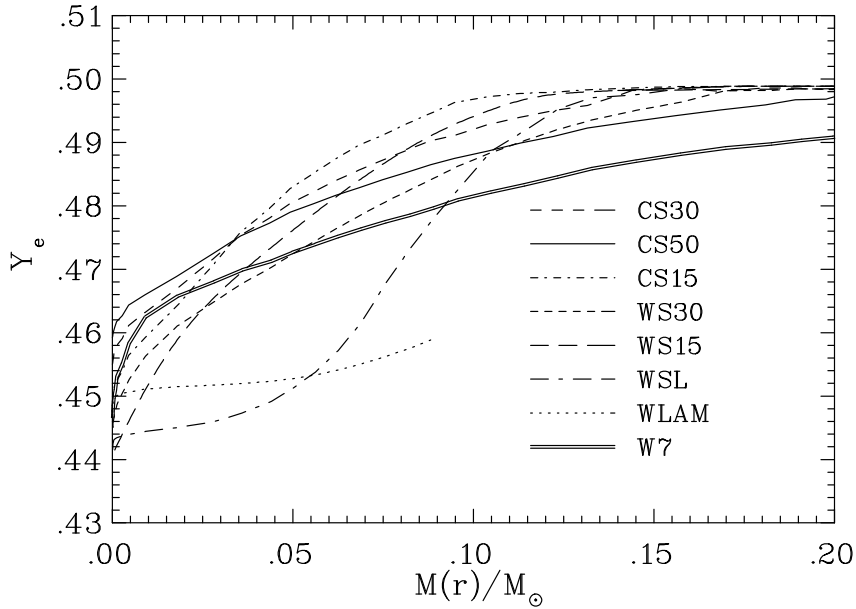}
\vspace{1cm}
\caption{$Y_e$, the total proton to nucleon ratio and thus a measure of 
electron captures on free protons and nuclei, after freeze-out of
nuclear reactions as a function of radial mass for different
models. In general it can be recognized that small burning front
velocities lead to steep $Y_e$-gradient which flatten with increasing
velocities (see the series of models CS15, CS30, and CS50 or WS15,
WS30, and W7).  Lower central ignition densities shift the curves up
(C vs. W), but the gradient is the same for the same propagation
speed. This only changes when the $Y_e$ attained via electron captures
during explosive burning is smaller than for stable Fe-group
nuclei. Then, $\beta^-$-decay during the expansion to smaller
temperatures and densities will reverse this effect (see models WSL
and WLAM discussed in more detail in the text).
\label{yemr}}
\end{figure}

	During the burning the central region undergoes electron
captures on free protons and iron peak nuclei. The central densities,
similar to W7 for both WS models but combined with higher
temperatures, result in more energetic Fermi distributions of
electrons and larger abundances of free protons. This leads to larger
amounts of electron captures on free protons and nuclei and smaller
central $Y_e$ values. Figure \ref{yemr} shows the final $Y_e$ during
charged-particle freeze-out and before long-term decay of unstable
burning products. Table 2 lists this value of $Y_{\rm e}$ in the
center, $Y_{\rm e,c}$, together with nucleosynthesis information,
which will be discussed later. In summary, $Y_{\rm e,c}$ is lower for
higher central densities and slower deflagrations, the latter leading
also to higher central temperatures. Figure \ref{yemr} and Table 2
also list two models WSL and WLAM that experience (partially or
entirely) deflagration velocities as small as the laminar speed, which
apparently do not follow this logic anymore. All models are discussed
in detail in the following paragraphs.


	WS15, corresponding to a slow deflagration with a
burning-front propagation of 1.5\% of the sound speed, reaches higher
central temperatures than WS30. The case WS30, corresponding to a
burning-front propagation of 3\% of the sound speed, has similar
densities, but slightly lower temperatures for each radial mass zone
in comparison to WS15 (also for a shorter duration). Besides the
different central value, both models also have a different central
$Y_e$ gradient (see Fig. \ref{yemr}). The WS15 curve, with lower
central values, reaches $Y_e$=0.4985 (inherited from the progenitor
white dwarf after He burning) at smaller radii than WS30. A smaller
deflagration speed causes a later arrival of the burning front at a
given mass coordinate.  Thus, matter at this mass coordinate has a
longer time to preexpand between the arrival of the central
information with sound speed and the arrival of the burning
front. Therefore, burning occurs there at a lower density (and
temperature) with smaller average electron energies, causing less
electron capture. This means that a smaller propagation speed produces
a smaller central $Y_e$, but reaches $Y_e$ = 0.4985 also at smaller
radii, i.e., produces a steeper $Y_e$ gradient. W7, with a different
description of the burning-front velocity, but on average a larger
speed, led to a higher central $Y_e$ and a flatter $Y_e$ gradient.

	CS15 and CS30, which also corresponds to a burning front
propagation with 1.5\% or 3\% of sound speed, behave in a fashion very
similar to WS15 and WS30 but are characterized by a smaller central
ignition density.  Because of a smaller ignition density the central
$Y_e$ values are larger. Otherwise the $Y_e$-gradients are the same
for CS15 and WS15 as well as CS30 and WS30, each pair having the same
burning-front speed. CS50, a case with 5\% of sound speed, has a more
extended region of decreased $Y_e$ out to larger masses (coming close
to the behavior of W7), but because of the lower ignition density and
lower central temperatures the central $Y_{\rm e,c}$ is larger again.



\begin{figure}[ht]
\epsfxsize=7cm
\epsfysize=6cm
\epsfbox[-865 76 -451 573]{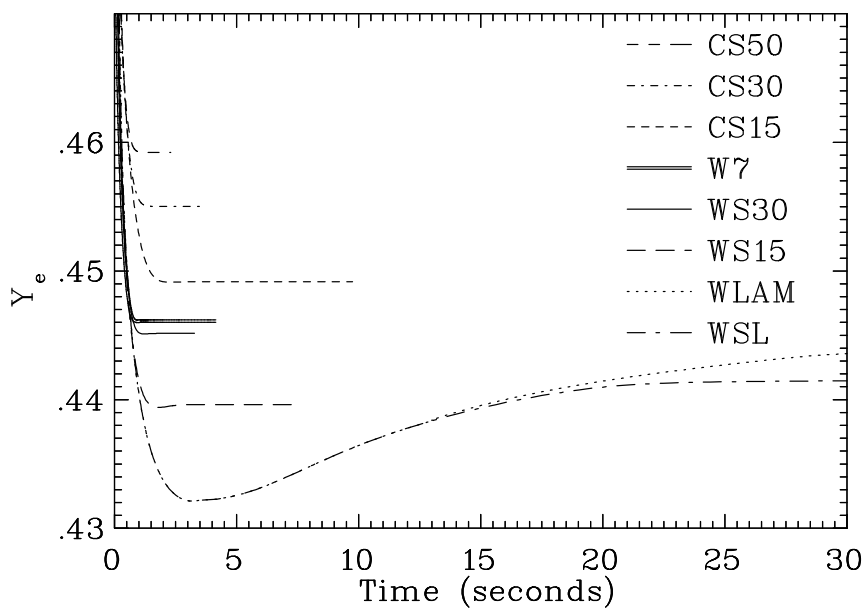}
\vspace{1cm}
\caption{
$Y_e(t)$ for the central mass zones in CS, WS, WSL, WSLAM and W7 
models. We see that $Y_e$ attains the lowest values for the highest ignition
densities and slowest burning front speeds. This tendency continues for the
extreme cases of laminar burning front models, but the slow expansion in
these models permits fast $\beta^-$-decays of short-lived nuclei in
NSE or QSE equilibria before charged-particle freeze-out.
\label{yetim}}
\end{figure}

	WSL is an additional case and corresponds to a calculation
with an initially laminar flame speed (out to 0.05~M${_\odot}$, see
Fig. \ref{yemr}), before an artificial acceleration is induced via
turbulent mixing of ashes. WSLAM starts in exactly the same manner;
however, the front stays laminar, i.e., only the minimum flame speed
is permitted. According to the previous discussion one would expect an
even steeper $Y_e$-gradient, and thus a lower central $Y_e$, owing to
higher temperatures during a longer duration time. The steeper
gradient can be seen outside 0.06 \ms~for WSL, where the front starts
to accelerate beyond the laminar speed. Inside the range of 0.06 \ms
the $Y_e$ is almost constant and larger than e.g., in WS15, which
experienced the same ignition density and larger flame speeds.  There
are two reasons for this behavior in the very central zones:

1. Initially during the burning, the lowest central $Y_e$-values of
$\sim$~0.432 are obtained as expected, but this value is -- in
contrast to all other cases encountered previously -- smaller (more
neutron-rich) than the Fe-group nuclei in the valley of stability (see
the most neutron-rich entries in Table 2, i.e., $^{50}$Ti, $^{54}$Cr,
$^{58}$Fe, and $^{64}$Ni with $Z/A$ values of 0.44, 0.444, 0.448, and
0.438). This leads to counterbalancing $\beta^-$-decays which win
against electron captures as densities and temperatures (slowly)
decrease while matter is still in an NSE. This causes an increase in
$Y_e$ to the displayed values before charged-particle freeze-out (see
Fig. \ref{yetim}).

\begin{figure}[hb]
\epsfxsize=7cm
\epsfysize=6cm
\epsfbox[-900 40 -495 614]{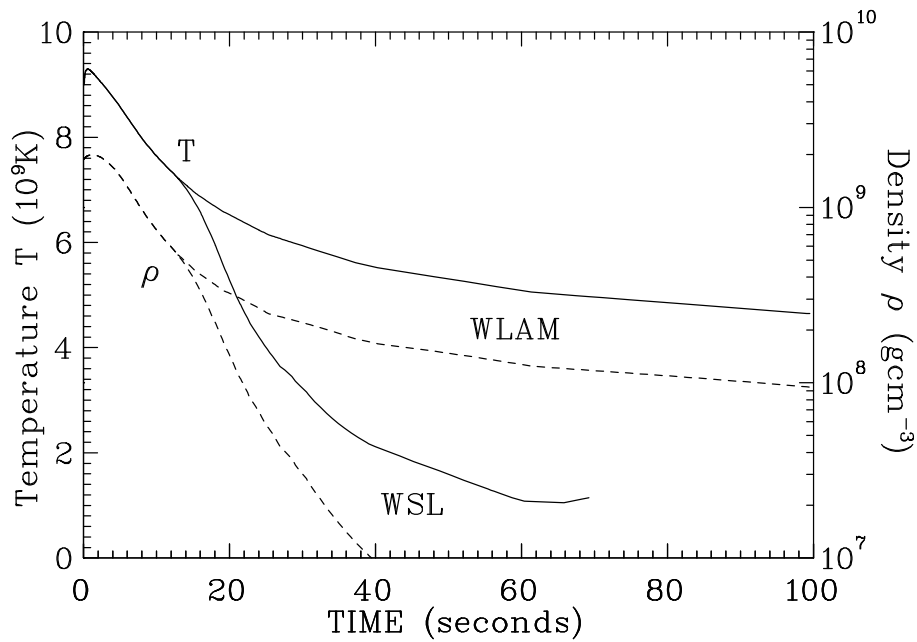}
\caption{
$\rho(t)$ and $T(t)$ for the central mass zones in WSL and WSLAM
models. The bifurctaion between WSL and WLAM marks the point where the
burning front was accelerated beyond the laminar speed in WSL,
resulting in a faster expansion.
\label{rttim}}
\end{figure}

2. During the laminar front propagation, the large amount of central
electron captures leads to neutrino losses which reduce the local
energy release to a point where the expansion is very small (see the
time dependence of $\rho$ and $T$ in Fig. \ref{rttim}). This keeps
temperatures and densities high for a prolonged period, and matter
stays in an NSE (the differences between WSL and WLAM emerge when the
burning front is accelerated in WSL beyond the laminar speed). This
NSE distribution of nuclei - with a total $Y_e$ more neutron-rich than
stable Fe-group nuclei - permits $\beta^-$-decay of short-lived nuclei
toward a total $Y_e$ corresponding to neutron-rich, stable Fe-group
muclei (beyond 0.44) as long as temperatures are high enough to ensure
an NSE. This results in the surprising fact that a burning front that
stays laminar (WLAM) causes a higher $Y_e$ after charged-particle
freeze-out than WSL (see changes in Fig. \ref{yetim} after 20~s).
Thus, a very small $Y_e$, which produces e.g., large amounts of
$^{48}$Ca ($Z/A$=0.417), can only be attained in models with a higher
ignition density, causing large amounts of electron capture, but with
nonlaminar burning fronts that permit a fast expansion rather than
keeping matter in NSE for a long time.

	With the exception of the laminar burning-front models just
discussed, which show a different behavior in Figure \ref{yemr}, we
can summarize the major options for slow deflagrations to change $Y_e$
values in the central part of SNe Ia explosions: (1) the burning-front
speed determines the $Y_e$ gradient, and the slowest speeds lead to
the smallest central values; (2) lower central ignition densities
cause larger $Y_e$ values, with the gradient, however, depending only
on the propagation speed. These features hold true as long as $Y_e$
values in explosive burning do not drop below 0.44, when competing
$\beta^-$-decays have also to be taken into account.  While the
correct conditions occurring in SNe Ia might be still forthcoming from
detailed multidimensional hydrodynamic calculations, this parameter
study shows how nucleosynthesis results can give important clues to
$v_{\rm def}$ and $\rho_{\rm c,ign}$.

Figures \ref{xi15c}ab, \ref{xi30c}ab, \ref{xiw7c}ab, and
\ref{xiwslc}ab show abundance plots (mostly of Fe-group nuclei) for
the central parts of CS15, WS15, CS30, WS30, CS50, W7, WSL, and WLAM
where electron capture plays an important role. We see that $Y_e$
values of 0.47-0.485 lead to dominant abundances of $^{54}$Fe and
$^{58}$Ni; values between 0.46 and 0.47 produce dominantly $^{56}$Fe;
values in the range of 0.45 and below are responsible for $^{58}$Fe,
$^{54}$Cr, $^{50}$Ti, and $^{62,64}$Ni; and values below 0.43-0.42 are
responsible for $^{48}$Ca.  Figure \ref{yemr} clearly indicates that
because of the flat $Y_e$-gradient of W7, the total amount of matter
experiencing the range of $Y_e$=0.47-0.485 is much larger than in
cases with slower deflagration speeds and larger $Y_e$-gradients. WS15
and CS15 have similar total amounts of $^{54}$Fe and $^{58}$Ni, but at
slightly different locations (see the $Y_e$-range in Fig. \ref{yemr}
and the different mass scales in Figs. \ref{xi15c}a and \ref{xi15c}b).
WS30 and CS30 contain a larger amount of $^{54}$Fe and $^{58}$Ni,
owing to the flatter $Y_e$-gradient.  CS50 comes close to the original
W7 model.





\subsection{Fe-Group Composition in Slow Deflagrations}

	One of the motivations for the present exercise is to get
constraints from comparison with solar Fe-group abundances. This can
provide a better understanding of the burning-front propagation and
test how the otherwise quite compelling features of the widely used
model W7 can be improved.  Figure \ref{w7sol} shows the ratio of
abundances produced in W7 to solar abundances.  These are the results
of recalculations of the original W7 with the present reaction rate
library and an increased accuracy in mass conservation in comparison
to earlier studies due to the screened NSE treatment at high
temperatures discussed in \S2.4.  Displayed are abundance ratios after
the decay of unstable nuclei, normalized to unity for $^{56}$Fe.  If
SN Ia events are a relatively homogeneous class, the comparison of
nucleosynthesis products with solar abundances is actually meaningful
without averaging over a complete sample.


	It is immediately obvious from Figure \ref{w7sol} that the
production of Fe-group nuclei in comparison to their solar values is a
factor of 2-3 larger than the production of intermediate nuclei from
Si to Ca.  When considering that SNe Ia produce about 0.8M$_\odot$ of
Fe-group nuclei in comparison to $\approx$0.1M$_\odot$ from SNe II,
and that the Ia/(II+Ibc) ratio is about 0.15-0.27 in our galaxy (van
den Bergh \& Tammann 1991; Cappelaro et al. 1997), SNe Ia are
responsible for more than 55\% of Fe-group nuclei.  Thus, even if an
isotope has no contribution from SNe II, this implies that the
isotopic ratios among the Fe-group in the SNe Ia ejecta should not
exceed the solar ratios by a factor of \(\sim 2\), in order to result
in solar ratios for SNe II + SNe Ia.  If we assume on the other hand
that an (Fe-group) isotope is made by SNe II in solar proportions,
then it has also to be made in solar proportions in SNe Ia to obtain
the solar mix for the sum of both contributions.  An overproduction by
SNe II would even ask for an underproduction in SNe Ia.  If we are
conservative and neglect the latter case, an overproduction of a
factor 1-2 in SNe Ia would be permitted. Multiplying this with an
average uncertainty factor of 2 would permit overproductions of 2-4,
dependent on the fact whether an SNe II contribution is existing or
not. In general we do not know this at present, and we take an
overproduction factor in SNe Ia of \(\sim 3\) as an alarm sign for
nucleosynthesis constraints.  In this respect, we notice in Figure 12
the overproduction of $^{54}$Cr and $^{58}$Ni by a factor of \(\sim
4-5\).  Here $^{54}$Cr is an $N=Z+6$ nucleus originating from the very
central regions with low $Y_e$, while $^{58}$Ni (and $^{54}$Fe) are
nuclei with $N=Z+2$ measuring the bulk neutron excess of the material
affected by the deflagration wave in the intermediate $Y_e$ range
$\sim 0.48$ (see also Table 2). Outside of 0.3M$_\odot$ in the
exploding white dwarf, where electron capture is not effective, the
neutron excess is only determined by the $^{22}$Ne ($N=Z+2$) admixture
to $^{12}$C and $^{16}$O in the original composition, stemming from
$^{14}$N in He burning, which in turn originated from all CNO-nuclei
in H burning.  The $Y_e$ or neutron-excess $\eta$ outside 0.3M$_\odot$
is thus a measure of the metal abundance (nuclei heavier than He) and
the galactic age of the white dwarf. The quoted calculations were
performed with $X(^{22}{\rm Ne}) = 0.025$, which corresponds to $\eta$
of \(\sim\) 30 \% higher than the solar metallicity, thus
overestimating the average value for "old" white dwarfs undergoing a
SN Ia event.


	Using an averaged metallicity of SN Ia progenitor systems less
than solar would reduce the overproduction of these nuclei. This is
shown in Figures \ref{w70sol} for a deflagration model like W7 but
with zero metallicity (W70). Figures \ref{w7full}a and 14b give the
comparison of the abundance distributions.  Bravo et al. (1992) have
performed a similar test. As the total SNe Ia contribution in our
galaxy is given by an integral over time or metallicity up to the
formation of the solar system, one expects average metallicities of
0.5-0.6 times solar. This brings the $^{54}$Fe/$^{56}$Fe,
$^{57}$Fe/$^{56}$Fe, $^{58}$Ni/$^{56}$Fe, and $^{62}$Ni/$^{56}$Fe
ratios within a factor of 2-3 of solar, respectively, which
corresponds to the present uncertainty range of thermonuclear reaction
rates and the minimum $\approx$50\% contribution of SNe Ia to the Fe
group. It does not reduce the $^{54}$Cr abundance sufficiently, which
originates solely from the central layers, where $Y_e$ is the smallest
and entirely due to electron captures rather than the metallicity in
terms of $^{22}$Ne.  An interesting aspect of the change in
metallicity, leading to reductions of $^{54}$Fe and $^{58}$Ni, is the
varying amount of early Fe ($^{54}$Fe before $^{56}$Ni decay) and late
Ni ($^{58}$Ni after $^{56}$Ni decay) in SN Ia ejecta, leading to
features that can be analyzed in observed spectra (see H\"oflich et
al.  1998).  This analysis covers one uncertainty regarding the
initial composition.  Another one would be a variation of the initial
$^{12}$C/$^{16}$O ratio, depending strongly on the initial white dwarf
mass and metallicity.  This has also been addressed recently by Umeda
et al. (1999) and H\"oflich et al.~(1999). Thus, while the metallicity
and initial composition are one set of parameters, the ignition
density and burning-front velocity represent another set, as outlined
in \S 2.  A burning front with a smaller velocity could reduce the
amount of material in the $Y_e$ range 0.47-0.485, where $^{54}$Fe,
$^{58}$Ni, and $^{62}$Ni are produced in large amounts (see e.g.,
Table 2).


	In Figures \ref{s15solw7}a and 15b, \ref{s30solw7}a and 16b,
\ref{s50solw7}, and \ref{slsolw7}a and 18b the ratios to solar
abundances (normalized to $^{56}$Fe) are displayed. Here the results
of the central slow deflagration studies have been merged with (fast
deflagration) W7 compositions for the outer layers, where $Y_e$ is
given by the initial $^{22}$Ne and not by electron captures. Thus,
these are not yet full delayed detonation models (which will be
discussed in the next subsection), but more preliminary approximations
to test the central Fe-group results. In comparison to Figure
\ref{w7sol} we see that in all cases the $^{58}$Ni problem is strongly
reduced and would be fully resolved when using also smaller
metallicities (see W7 vs. W70). This is due to the steeper $Y_e$
gradient which produces less matter in the intermediate $Y_e$ range
0.47-0.485. The smaller propagation speed has, however, also the
consequence that $Y_e$ dips deeper in the central layers of WS15 and
WS30 than in W7. Such $Y_e$'s lead to the overproduction of $^{50}$Ti
and $^{54}$Cr. This is not the case for CS15 and CS30, due to the
lower ignition density.

	Another interesting point surfaces, which was also addressed
preliminarily in Thielemann et al. (1996) and also Meyer, Krishnan, \&
Clayton (1996) and Woosley (1997b). If one takes the results of
presently existing and still crude SNe II nucleosynthesis calculations
from initiated explosions as well as the results of W7, it turns out
that for some intermediate mass and all Fe-group elements the most
neutron-rich nuclei are drastically underproduced. The central $Y_e$
values of slow deflagration models comes close to conditions, where
these nuclei are produced in a normal freeze-out. This has the effect
that nuclei like $^{50}$Ti, $^{54}$Cr, $^{58}$Fe and partially
$^{64}$Ni or $^{48}$Ca are produced for $Y_e$ values below 0.46 (or
$\eta=1-2Y_e=0.1$). Whether this leads after integration over all mass
zones just to solar abundances or to a strong overproduction will be
tested here (see also Table 2).  Khokhlov et al. (1992) undertook
already a preliminary assessment of this question but did not consider
all of these nuclei in their calculations. Woosley (1997b) tested
models with higher ignition densities and burning-front velocities,
whether such elements can be produced.  Here we take again the
philosophy to obtain constraints for the "average" case of SNe Ia, to
test on the one hand whether they can produce such isotopes at all,
and if so, whether constraints can be set for the conditions in order
to avoid overproductions beyond the often mentioned factor of 3.





	It is a well-known fact that SNe Ia are not identical (see
e.g., Hamuy et al. 1995) and that therefore a continuous superposition
of such models has to be responsible for the solar Fe-group
composition. On the other hand, we also know that the majority of SNe
Ia come from a narrow window of conditions (Branch 1998), i.e., the
notion of an average SN Ia event makes sense.  If Fe-group elements
are produced by SNe Ia to $\sim$ 55 \% and Ia's were the only sources
of the neutron-rich Fe-group nuclei, we must conclude that an
overproduction of a factor of 3 is permitted and that a larger
overproduction has to be avoided for the average event.  We see that
for the cases discussed before (CS15, WS15, CS30, WS30) $^{54}$Fe and
$^{58}$Ni can be produced within the permitted uncertainty limits
($^{58}$Ni however having a tendency for overpoduction). These are the
nuclei with intermediate $Y_e$ (0.47-0.485), which measure the $Y_e$
gradient via the amount of mass contained in that interval. CS50,
which has a larger burning-front speed and a flatter slope, starts to
overproduce these nuclei.

	The more neutron-rich Fe-group nuclei depend more on the
central layers, which experience lower $Y_e$-values.  $^{54}$Cr is
produced within permitted limits in CS15 and CS30 but is overproduced
in WS15 and WS30 owing to the higher ignition densities and central
$Y_e$ values that are too low. $^{50}$Ti is produced close to solar
values for CS15, underproduced by CS30, well produced for WS30, and
clearly overproduced in WS15, preferring the central $Y_e$ values of
CS15 and WS30, which are similar.  $^{64}$Ni is essentially only made
in large quantities for such low-$Y_e$ conditions as in WS15. Close to
solar values cannot be attained for $^{48}$Ca in any of these models,
and its production would require lower $Y_e$ values.  An attempt to do
this was a purely laminar, i.e., the slowest possible, burning
front. The display in Figure \ref{slsolw7}a and 18b makes clear that
this is not possible, as discussed in \S3.1. The main reason is that
such conditions produce nuclei which are unstable against
$\beta^-$-decay. In intermediate phases sufficiently low $Y_e$ values
are attained as a result of electron captures. NSE or
Quasi-statistical equilibrium (QSE) redistributes abundances according
to the then obtained smaller $Y_e$ values. $\beta^-$-decay of the
short-lived neutron-rich isotopes leads to an increase in $Y_e$ during
the expansion, when the densities and temperatures (and therefore the
electron capture rates) decrease, before freeze-out from charged
particle reactions and NSE/QSE (see Fig. \ref{yetim}).  Thus, to
produce a nucleus like $^{48}$Ca in sufficient amounts, only very
high-density ignitions (followed by a fast expansion to avoid the
influence of $\beta^-$-decays) of progenitor systems, which barely
avoid an accretion induced collapse (AIC), might be responsible (see,
e.g., Nomoto \& Kondo 1991; Woosley 1997b).

	From this exercise we see that not all of these neutron-rich
nuclei can be made in similar proportions for one set of deflagration
parameters unless a detailed fine-tuning of $v_{\rm def}(r)$ and
$Y_e(r)$ is performed or multidimensional propagation of the burning
front produces exactly a superposition of conditions as needed to fit
all abundance constraints.  In general CS15 and CS30 seem to be better
models than WS15 and WS30 in terms of avoiding a large overproduction
of neutron-rich elements. The burning-front speed in the central
layers seems to be constrained to values below 5\% of the sound speed
in order to avoid the overproduction of $^{58}$Ni (see CS50 and Figure
\ref{s50solw7} as well as Table 2 with comparable problems found for
W7 and W70).  If on the other hand there were no other sites than SNe
Ia to produce isotopes like $^{50}$Ti, $^{54}$Cr, $^{58}$Fe, $^{64}$Ni
or $^{48}$Ca, we would have to overproduce in comparison to solar by
about a factor of 2, and some features of the models WS15, WS30 and
even higher density events were needed, which can fill in the
remaining deficiencies. Thus, one would need the majority of events
similar to CS15 or CS30 and a smaller number of higher density
ignitions to produce the more neutron-rich nuclei. This could
guarantee on the one hand some overproduction, which (in combination
with SNe II) would permit solar abundances in total, and avoid on the
other hand not-permitted overproduction.

\subsection {Delayed Detonation}

	The final aim, after constraining the central slow
deflagration part of SNe Ia, is to find also composition constraints
for the deflagration detonation transition (DDT). As discussed in \S
2.3, we chose transition densities of 3.0, 2.2, and 1.7$\times
10^7$~g~cm$^{-3}$ for the given models WS15 and CS15, which turn them
into WS15DD3, WS15DD2, WS15DD1 or CS15DD2 and CS15DD1.  Figures
\ref{wdd12} - \ref{cdd12} show the abundance distributions of slow
deflagrations combined with delayed detonation models against the
expansion velocity and $M_r$ of DD models. The central regions of
these models have been shown before in Figures \ref{xi15c} and
\ref{xi30c}, thus the abundance distributions of neutron-rich species
such as $^{54}$Cr, $^{50}$Ti, $^{58}$Fe, and $^{62}$Ni are not
repeated here.

	As the deflagration wave propagates outward, the white dwarf
gradually expands to undergo less electron capture and thus mostly
\ni~ is synthesized.  Eventually, the deflagration enters the region
of incomplete Si burning and explosive O-Ne-C-burning, where the
transition to a detonation occurs. For comparison the abundance
distributions of W7 and W70 were shown in Fugure \ref{w7full}.  The
total masses of $^{56}$Ni($^{56}$Fe) produced in these combined models
have been summarized in Table 1.




	These theoretical abundance distributions can be compared with
the observed expansion velocities of several elements as estimated
from supernova spectra.  It is seen that WDD2 and WDD1 produce two
Si-S-Ar peaks at low velocity ($\sim$ 4,000 \kms) and high velocities
(10,000 - 15,000 km s$^{-1}$).  The intermediate mass elements at low
velocities are important to observe at late times in order to
distinguish between models.  In particular, the minimum velocity of Ca
in WDD models is $\sim$ 4,000 \kms, which would be higher for a faster
deflagration.

	The Ca velocities should be compared with the observed minimum
velocities of Ca indicated by the red edge of the Ca II H and K
absorption blend (Fisher \etal 1995).  The lowest velocities of O and
Mg also provide interesting constraints.  For example, SNe 1990N,
1992A, and 1991T show O in the wide velocity range from $\approx$
10,000 to 20,000 \kms\ (Leibundgut \etal 1991a; Jeffery \etal 1992;
Mazzali \etal 1993; Kirshner \etal 1993).  For W7 and WDDs, the
minimum O velocity is 12,000 - 15,000 \kms.  The observed O velocity
as low as 10,000 \kms~ may indicate a mixing of O in the velocity
space.

	Meikle \etal (1996) have observed a P Cyg-like feature at
$\sim$ 1.05/1.08 $\mu$m in SN 1994D and 1991T.  They note that, if
this feature is due to He, He in SN 1994D is likely to be formed in an
alpha-rich freeze-out and mixed out to the high-velocity layers
($\sim$ 12,000 \kms).  The maximum velocity of He is 5,000 - 6,000
\kms~ in WDDs, being slower than $\sim$ 9,000 \kms~ in W7, so that
more extensive mixing of He would be required for WDDs than in W7.
(Note that $^4$He is plotted neither in Fig. \ref{w7full} nor in Figs. 
\ref{wdd12} - \ref{cdd12}, but its location can be easily identified
with the region where the $^{58}$Ni abundance forms a plateau up to
the point where it turns over to $^{54}$Fe on the same level. This is
the transition from alpha-rich freeze-out to incomplete Si burning.)
Alternatively, if the feature is due to Mg, the Mg velocity is
confined to 12,500 - 16,000 \kms~ in SN 1994D, which is consistent
with W7 (13,000 - 15,000 \kms).  For WDDs, on the other hand, the
minimum velocities of Mg are 14,500 \kms~ (WDD1), 16,500 \kms~ (WDD2),
and 18,000 \kms~ (WDD3), and the latter two models seem to have too
high velocities.

	Mixing of ejected material in velocity space could occur
convectively during the propagation of the deflagration wave (Livne
1993).  Nonspherical explosions induced by delayed detonations could
also produce nonspherical abundance distribution, i.e., elemental
mixing in the velocity space. From the calculated synthetic spectra
and their comparison with the observations, more advanced methods
(Harkness 1991) do not favor mixing opposite to initial suggestion by
Branch \etal (1985).

\section {Yields of SNe Ia and Galactic Chemical Evolution}

\subsection{Features of SN Ia Nucleosynthesis}

	Complete isotopic compositions of WDD and CDD models are given
in Table 3. Table 3 assumes full decay of all unstable species. We
provide separately in Table 4 abundances of long-lived radioactive
nuclei, of importance either for gamma-ray detection, extinct
radioactivities, or chemical evolution. The abundances are compared
with solar abundances in Figures \ref{csdsol} - \ref{wsd3sol}, which
are normalized to $^{56}$Fe. These Figures complement the earlier
Figures \ref{s15solw7} - \ref{slsolw7}, where the fast deflagration W7
was utilized in the outer layers rather than delayed detonation
models. The major conclusions on the Fe-group composition remain the
same as discussed before, in \S 3.1, owing to the fact that they are
given by the central slow deflagrations. What can be studied here is
the additional variation in the ratio of Fe-group to intermediate mass
nuclei, which depends on the deflagration-detonation transition. Of
course the variation in $^{56}$Fe (originating from $^{56}$Ni) in the
outer detonation layers also influences the ratio of neutron-rich
Fe-group isotopes (from central locations) to $^{56}$Fe.  Table 3
includes besides all DD-models (CDD1 short for CS15DD1 etc.)  also W7
and W70 updated with the latest reaction rate set and improved
accuracy by using a screened NSE treatment for long duration times at
temperatures beyond $6\times 10^9$~K (Hix \& Thielemann 1996).  The
ratios to solar abundances for W7 and W70 were shown in Figures
\ref{w7sol} and \ref{w70sol}.  The essential features due to the DDT,
as displayed in Figures \ref{csdsol} - \ref{wsd3sol} and in Table 3
can be summarized as follows:




1. The synthesized amount of Fe and thus the ratio between the
intermediate mass elements and Fe, Si-Ca/Fe, is sensitive to the
transition density from deflagration to detonation, as was already
shown in Table 1. Among the WDD models, WS15DD2 produces only about
25\% more \ni~ than W7 ($\sim$ 0.6 \ms) but more Si-Ca than W7 by 40\%
(Fig. \ref{wsdsol}b), since more oxygen is burned in the outer layers.
Therefore, the Si-Ca/Fe ratios are moved up to a certain extent.
WS15DD1 has even larger Si-Ca/Fe ratios, which are close to solar
ratios (Fig. \ref{wsdsol}a).  This is not indicated for SNe Ia, owing
to observations of low-metallicity stars reflecting the average SN II
behavior (a reasoning outlined below in more detail). However, direct
observations of the Si-Ca/Fe ratio in SNe Ia remnants (Tycho, SN 1006,
etc.) are highly important and needed in order to distinguish between
the models (Hughes et al. 1995; Miyata et al. 1998; Hwang, Hughes \&
Petre 1998).

2. Neutron-rich species such as $^{54}$Cr and $^{50}$Ti are mostly
produced in the slow deflagration phase. The degree of their
overproduction with respect to $^{56}$Fe depends also on the mass of
$^{56}$Ni produced in the outer detonation layers, as seen in Figure
\ref{wsd3sol}.  This also explains why in Table 2 the entries for the
same central models (WS15, CS15 etc.) change for neutron-rich species,
although the central part from which these neutron-rich species
originate is unaffected by the detonation. The reason is that the
ratios in comparison to $^{56}$Fe are taken.

3. There are some Fe-group contributions from alpha-rich freeze-out
and incomplete Si-burning layers that depend on the DDT. Figures
\ref{wdd12} - \ref{cdd12} show that the mass region experiencing
incomplete Si burning (indicated by the $^{54}$Fe plataeu) decreases
in the sequence DD1-DD3. The region indicated by the $^{58}$Ni plateau
experiences alpha-rich freeze-out (the He abundance is not shown here)
and increases in the sequence DD1-DD3. $^{52}$Fe (decaying to the
dominant Cr isotope $^{52}$Cr) and $^{55}$Co (decaying to the only
stable Mn isotope $^{55}$Mn) are typical features of incomplete Si
burning. $^{59}$Cu (decaying to the only stable Co isotope $^{59}$Co)
is a typical feature of an alpha-rich freeze-out.  For these reasons
we see the strongest $^{52}$Cr and $^{55}$Mn overabundances in DD1 and
the strongest appearance of $^{59}$Co (while still underabundant) in
DD3.

\subsection{The Role of SN Ia and SN II Contributions}

	The chemical evolution of galaxies is dominated by its main
contributors SNe II, SNe Ia, and planetary nebulae. The latter do not
contribute to the element abundances in the range O through Ni
(although a few specific minor isotopes can be produced in the
s-process). Thus for the aspects considered here, we have to explain
galactic evolution and also solar abundances by the combined action of
SNe II and Ia. The ratio of Fe-group elements to Si-Ca in SNe Ia is of
specific importance, in order to see how the overabundance of O-Ca/Fe
(in comparison to solar) in SNe II can be compensated. Combined
nucleosynthesis products of SNe Ia and SNe II with varying ratios can
be compared to solar abundances. An important aspect for such an
undertaking is, however, to test the individual components against
existing observations first, before trying to attain a good solar mix
(with possibly wrong predictions for the individual SN I and SN II
components).

	Although we do not have a good quantitative measure from SN Ia
observations for the Fe-group to Si-Ca ratio, the fast deflagration
model W7 seems to give a good overall agreement via synthetic spectra
calculations with observed Ia spectra (see Branch 1998 for a
review). W7 has specific deficiencies in the global isotopic Fe-group
composition from the inner layers, as discussed before, e.g., with
respect to $^{58}$Ni and $^{54}$Cr, but this does not affect the
element ratios too strongly.

	A similar or even worse situation is found for the
quantitative analysis of SN II spectra. However, an independent tool
exists that measures the integrated SN II yields: observed surface
abundances of low-metallicity stars, which witness the abundances of
the interstellar medium at their point of formation during early
galactic evolution, when only SNe II contributed. For metallicities in
the range $-2<$[Fe/H]$<-1$ we expect the integrated (mass averaged)
properties of SNe II (see e.g., Nakamura et al. 1999). Here [x/y] is
defined as log$_{10}$[(x/y)/(x/y)$_\odot$]. Such abundance features
were reproduced with nucleosynthesis products of SNe II as a function
of stellar mass, taken from the calculations by Nomoto \& Hashimoto
(1988), Hashimoto \etal (1996), and Thielemann \etal (1996) as
summarized in Hashimoto (1995), Tsujimoto \etal (1995), and Nomoto et
al. (1997a). SNe II yields, integrated from $m_l$ = 10 \ms~ to $m_u$ =
50 \ms~ with a Salpeter IMF, are also given in Table 3.  The upper
mass bound $m_u$ is chosen to reproduce [O/Fe] = +0.4, which is
consistent with the observations of low metallicity stars for [Fe/H]
$<$ -- 1. Such observations in low-metallicity stars give also the
best constraints on average Fe-group abundances of SNe II, which are
poorly known theoretically owing to the still existing lack of
self-consistent core collapse supernova models. The representation of
the Fe-group (beyond Ti) closest to the low metallicity observations
seems to be the one of our 20\ms~star (see Figure 5b in Thielemann
\etal 1996 and note that the observed Co abundance has come down to
about -0.1, i.e., it shows a better agreement with the dashed line).

	A possible deviation from solar ratios has to be made up by an
opposite behavior of SNe Ia setting in at about [Fe/H]=$-$1 in order
to attain solar values at [Fe/H]=0. Fe-group elements for which
information is available are Ti, Sc, Cr, Mn, Co, Ni (Magain 1987,
1989; Gratton \& Sneden 1988, 1991; Gratton 1989; Zhao \& Magain 1990;
Nissen et al. 1994). They lead to typical uncertainties of 0.1 dex and
one finds average SN II values of 0.25, 0, -0.1, -0.3, -0.1, -0.1.
Taken the typical uncertainty of 0.1 dex, this leaves Ti and Mn as
elements with clear signatures for a SN II behavior different from
solar, which asks for the opposite SN Ia behavior.  Cu and Zn start
having strong s-process contributions. We avoid their discussion
because of these complications and because they are not a clear
indication for the required SN Ia signature.

	Ti is dominated by the isotope $^{48}$Ti (see Table 3), i.e.,
we have to relate the element ratio Ti/Fe to $^{48}$Ti/$^{56}$Fe.  We
see that for all models that underproduce Si-Ca (as needed for SNe
Ia), i.e., the DD2, DD3 and W7 models, $^{48}$Ti is also underproduced
by similar amounts. This agrees with the observational trend that Ti
is dominantly produced by SNe II and can be understood from the
$^{48}$Cr abundances (decaying to the main Ti isoptope
$^{48}$Ti). $^{48}$Ti is only produced in a strong alpha-rich
freeze-out as it occurs in SNe II. The small region of a weak
alpha-rich freeze-out, indicated in Figure \ref{rhoT}, is not
sufficient.

	The tendency is not so clear for Mn. The only stable Mn
isotope is $^{55}$Mn, typically produced from unstable $^{55}$Co. This
isotope results (1) from incomplete Si burning with a relatively high
$Y_e$ (e.g., 0.4985; see its production in SNe II in Fig. 1 in
Thielemann et al. 1996 and the present Figs \ref{wdd12} - \ref{cdd12}
in the incomplete Si-burning regions) or (2 from a somewhat reduced
$Y_e$ (around 0.49) in complete Si burning (see Figs. \ref{xi15c} -
\ref{xi30c}). This can be understood within the framework of
quasi-equilibrium groups in Si burning, in our case the Si and the Fe
group.  Incomplete burning leads to a small total abundance in the Fe
group in comparison to the Si group. Hix \& Thielemann (1996) found
that in such a case the composition in the Fe group is typically more
neutron-rich than expected from the global $Y_e$. Hence, $^{55}$Co
with a $Z/A$=0.49 is in incomplete burning also produced for
conditions with a global $Y_e$ of 0.4985.  Both locations (in complete
and incomplete burning) can be nicely seen in Figure \ref{w7full},
with the additional $Y_e$ information taken from Figure \ref{yemr}. As
discussed before, the same is found in the central parts in Figures
\ref{xi15c} - \ref{xi30c} and globally in Figures \ref{wdd12} -
\ref{cdd12}.  The problem is, however, that with the exception of the
DD1 models and W7, all DD models underproduce Mn/Fe in comparison to
solar, opposite to what is required from average SN II yields inferred
from low-metallicity observations. It appears that the dominant source
for Mn is the incomplete Si-burning region, which is most extended in
the DD1 models.

	$^{52}$Cr is the dominant nucleus of the Cr isotopes. DD2
shows a slight overabundance, DD1 a stronger overabundance. $^{52}$Cr
originating from $^{52}$Fe is also a nucleus dominated by incomplete
Si burning and therefore this behavior is understandable. The light
underabundance in SNe II can thus be compensated by SNe Ia.  Co from
SNe II is slightly underproduced. All our calculations show an
underabundance in SN Ia models. It seems that the alpha-rich
freeze-out nucleus $^{59}$Cu (decaying to the only stable Co isotope
$^{59}$Co) is never produced sufficiently. However, there exists the
same problem in SNe II (Nakamura et al. 1999) and possibly other
(nuclear?) sources might be the origin of this behavior.  $^{58}$Ni
originates from neutron-rich central regions and partially from
alpha-rich freeze-out (where also $^{62}$Ni is produced via $^{62}$Zn
decay).  We have worked hard on our models to avoid an overproduction. 
Thus a slight underproduction in SNe II can easily be compensated.
One should also have a look at Bravo et al. (1993) and Matteucci et
al. (1993), who performed chemical evolution calculations and
addressed some of these questions with the then existing observational
and model constraints.

	We finally want to return to neutron-rich Fe-group isotopes
and point to a different witness of galactic evolution.  While
typically astronomical observations can give information only about
element abundances (with a few exceptions from molecular lines in
stars), there exists one source of isotopic information, the so-called
isotopic anomalies in meteorites. They are usually contained in
"inclusions" of primitive meteorites consisting of minerals with much
higher melting temperatures than the surrounding matter. This gives
some indication that they originate from unprocessed "star dust" that
survived temperatures in the early solar system and can give direct
clues about its stellar origin. This is essentially proven for some
SiC grains, graphites and diamonds (e.g., Zinner, Tang, \& Andrews
1989; Zinner 1995; Travaglio et al. 1999).

	The nuclei $^{48}$Ca, $^{50}$Ti, $^{54}$Cr, and $^{58}$Fe,
which play a crucial role in the central layers of SN Ia explosions,
can also be found in isotopic anomalies (Lee 1988; V\"olkening \&
Papanastassiou 1989, 1990; Loss \& Lugmair 1990). They occur, however,
only in Ca-Al-rich inclusions whose history is less clear and some
chemical processing in the early solar nebula has probably happened.
Nevertheless, it is interesting to investigate whether these observed
anomalies can lead to any indication of their origin. Harper et
al. (1990) found a correlation with the r-process nucleus $^{96}$Zn.
This would have pointed toward a SN II origin if the r-process
originates from SNe II. On the other hand, Ireland (1994) found a
probe where no correlation with r-process anomalies existed. Thus, no
very clear indication for their SN II or SN Ia origin (or both?) 
presently exists. In the present paper, however, we applied the
working hypothesis that their source is given by SN Ia ejecta, which
is also indicated by the $Y_e$-constraints from SN II abundance
observations (Thielemann et al. 1996).  In case of an additional SN II
contribution, even more stringent limits would exist.

\subsection{Ratios of SN Ia to SN II Events in the Galaxy}

	We have full results for nucleosynthesis products of SNe Ia
from the four models, WS15DD1-3, CS15DD1-2, W7, and W70, and the
results are shown in Figures \ref{csdsol} - \ref{wsd3sol} and
\ref{w7sol} - \ref{w70sol}. CS30 and 50 as well as WS30 were
calculated only for the central slow deflagration layers and not full
delayed detonation models. This, as it affects the composition within
the Fe group, might not be a sufficient quantitative basis with
respect to relative abundances within the Fe group. But as the central
deflagration speeds have minor influence on the total abundance of
$^{56}$Ni ($^{56}$Fe), the different DD models are probably sufficient
to determine the best Si-Ca/Fe ratios.  The observational data by
Gratton \& Sneden (1991) and Nissen et al. (1994) for [x/Fe] at low
metallicities ($-2<$[Fe/H]$-<1$), x standing for elements from O
through Ca, show an enhancement of the alpha elements (O through Ca)
by a factor of 2-3 (0.3 to 0.5 dex in [x/Fe]) in comparison to
Fe. This is the clear fingerprint of the exclusive contribution of SNe
II in early galactic evolution.  It has long been the aim of chemical
evolution calculations to explain this behavior, among the most recent
ones being Tsujimoto et al. (1995), Timmes, Woosley, \& Weaver (1995),
Pagel \& Tautvaisine (1995, 1997), and Kobayashi et al. (1998).

	Tsujimoto et al. (1995) tried to determine the ratio of SN Ia
to SNe Ib+II events in the Galaxy by aiming for a best fit to an
overal solar abundance pattern from O to Ni. They made use of the
mass-averaged SN II yields from 10 to 50 M$_\odot$ as shown in Table 3
and earlier results for W7. With the aid of a chemical evolution model
they obtained a ratio $N_{\rm Ia}/N_{\rm II}=0.12$ of the total number
of SNe Ia to SNe II (+Ib) that occurred in our Galaxy.  This resulted
from an overal best fit to the observed abundances and is consistent
with the fact that the observed estimate of the SNe Ia frequency is as
low as 10 \% of the total supernova rate. The observational
constraints for this ratio range from values around 0.15 (van den
Bergh \& Tammann 1991) to more recent determinations of 0.27 with
errors of almost a factor of 2 (Cappelaro et al. 1997).

	Rather than performing a full galactic evolution model, we
want to concentrate on a typical abundance ratio, Si/Fe, and its
contribution from SNe II and Ia. The averaged SNe II yields from our
theoretical models summarized in Table 3 correspond to a
[Si/Fe]=0.347.  The results of the 20 M$_\odot$ model of Thielemann et
al. (1996) would correspond to [Si/Fe]=0.304, which is close to the
value of Gratton \& Sneden (1991) and Ryan, Norris, \& Beers (1996) of
0.3. One should, however, be aware of the typical uncertainties of
0.1~dex.  Varying the SN Ia models among the above given list (W7,
W70, DD1, DD2, and DD3) and making use of Si and Fe from the averaged
SNe II ejecta or the 20 M$_\odot$ star, leads also to a required
Ia/Ib+II ratio in order to obtain the solar mix of Si/Fe with both
contributions. This ratio is derived in the following way:

\begin{eqnarray}
   \left({M({\rm Si})\over M({\rm Fe})}\right)_\odot & = 
   &{N_{\rm Ia} M_{\rm Ia}({\rm Si}) + N_{\rm II} M_{\rm II}({\rm Si})}\over 
   {N_{\rm Ia} M_{\rm Ia}({\rm Fe}) + N_{\rm II} M_{\rm II}({\rm Fe})},\\
  {N_{\rm Ia}\over N_{\rm II}}& = 
  & {{M_{\rm II}({\rm Si}) - M({\rm Si})/M({\rm Fe})_\odot M_{\rm II}
({\rm Fe})} \over {M({\rm Si})/M({\rm Fe})_\odot M_{\rm Ia}({\rm Fe}) 
- M_{\rm Ia}({\rm Si})}}. 
\end{eqnarray}

	The results for the obtained ratios are shown in Table 5.  It
is immediatey obvious that the DD1 models do not result in a physical
solution. With their Si-Ca/Fe ratios being almost solar, they cannot
compensate for the SNe II contribution, which have larger than solar
values. There is a clear need for SNe Ia to produce ejecta with
Si-Ca/Fe less than solar. Otherwise all models seem to be covered by
the error bars for the Ia/(Ib+II) ratios laid out by van den Bergh \&
Tammann (1991) and Cappelaro et al. (1997). The DD3 values, especially
for the 20~M$_\odot$ Si/Fe ratios being closest to the low-metallicity
[Si/Fe], might somewhat indicate the lower limit. It should be noted
that although the Si/Fe ratios in the DD1 - DD3 models show a
monotonous behavior (i.e., decline), this does not lead to a
monotonous variation in Ia/(Ib+II) ratios, owing to the form of
equation (4), where differences and not the Si/Fe ratios enter. This
results in negative values for DD1 (due to a negative denominator) and
goes through a maximum (when the denominator starts to be positive but
is small) before decreasing to the DD3 values with an increasing
denominator. This is due to the fact that Si masses decrease with the
DD1 - DD3 sequence, while Fe increases.

	With a low SN Ia frequency of Ia/(II+Ib)=0.15 and W7
abundances, $^{56}$Fe from SNe Ia amounts to about 55\% of total
$^{56}$Fe (see the discussion in \S 3.2). Larger values would increase
this contribution.  When assuming that SNe II do not produce
neutron-rich species like $^{54}$Cr and $^{50}$Ti at all, we permitted
for the abundance ratios between such nuclei and $^{56}$Fe an
overproduction factor of $\sim$ 4. With the larger SN Ia frequencies
of Table 5 and Cappelaro et al. (1997) such permitted overproductions
would need to be further reduced and more stringent limits would
apply.

\section{Summary and Outlook}

	From the very early days of explosive nucleosynthesis
calculations, when no direct connection to astrophysical sites was
possible yet, it was noticed (Trimble 1975) that the solar Fe-group
composition could be reproduced with a superposition of matter from
explosive Si burning with about 90\% originating from a $Y_e$=0.499
source and 10\% from a $Y_e$=0.46 source.  We have discussed in some
detail in the present paper that the central part of SNe Ia could be
this second source, while Thielemann et al. (1996) have shown that SN
II ejecta with $Y_e$$<$0.498 could cause serious problems.

	New results by Hachisu \etal (1999a, 1999b; see also Li \& van
den Heuvel 1997), which include wind losses in the interaction of
binary systems, come to the conclusion that the majority of SNe Ia
progenitor systems experience hydrogen accretion on white dwarfs at a
rate that has them grow toward the Chandrasekhar mass through steady
H- and He-burning. This leads to the single degenerate Chandrasekhar
mass scenario (SD/Ch). Binary systems with steady H-burning on
accreting white dwarfs and effective accretion rates as high as $\dot
M >$ 1 $\times$ 10$^{-7}$ M$_\odot$ yr$^{-1}$ might correspond to
observed supersoft X-ray sources. This also would lead to low central
ignition densities of the Chandrasekhar mass white dwarf at the
thermonuclear runaway ($<$ 2 \e9 \gmc), which correspond to the C
series of the models discussed in the present paper. A small fraction
can deviate from steady H burning and would experience weak hydrogen
flashes near the end of the accretion history. Such cases would
correspond to our W series of models and even higher ignition
densities.  Kobayashi et al. (1998) discussed metallicity effects and
their influence on the delay time for the appearance of SNe Ia in
galactic evolution.

	Our nucleosynthesis results of the present paper are quite
consistent with these scenarios and favor case C. The reason is that
too high ignition densities lead to a high degree of electron captures
and small $Y_e$ values, which would cause an overproduction of
$^{54}$Cr and $^{50}$Ti in excess of what is permitted for SNe Ia in a
solar mix. We have to make two reservations here. New shell model
calculations (Dean et al. 1998; Caurier et al. 1999) indicate that the
electron capture rates could be substantially reduced in comparison to
the rates by Fuller et al. (1980,1982,1985) employed here. The
application of these new capture rates might also permit models of our
W series without serious deviations from the allowed central $Y_e$
values.  In addition, such conclusions deal only with the dominant or
average SN Ia events. If more neutron-rich nuclei like $^{48}$Ca are
also the result of SN Ia nucleosynthesis, an occasional event with a
significantly higher ignition density is required, a fact also
consistent with the above scenario.
 
	Our nucleosynthesis results imply that for the dominant events
the central density of the Chandrasekhar mass white dwarf at
thermonuclear runaway must be as low as or lower than \ltsim 2 \e9
\gmc, though the exact constraint depends somewhat on the flame speed. 
Here is the point where nucleosynthesis predictions can also be of
help for providing constraints to the supernova modeling and the
burning-front velocity.  A carbon deflagration wave, propagating as
slow as $v_{\rm def}/v_{\rm s} \sim 0.015$ or even slightly slower,
would be the ideal choice for the neutron-rich species such as
$^{54}$Cr and $^{50}$Ti; see cases CS15 and WS30. The latter case
makes also clear that slightly larger ignition densities than 1.5 \e9
\gmc~(case C) are permitted if the flame speed is increased
appropriately.  An acceleration of the flame speed in the outer part
of the central layers is permitted to about 3\% of the sound speed or
slightly more, but should clearly stay below 5\%. For such fast
burning fronts the $Y_e$-gradient becomes very flat and too much
material in the range $Y_e$ = 0.47-0.485 is produced. This leads to
dominant abundances of $^{54}$Fe and $^{58}$Ni, a feature that was
already prominent in W7, and causes an excessive overproduction of
$^{58}$Ni in galactic evolution.  Our calculations were performed with
a constant fraction of the sound speed.  An acceleration of the
burning front is expected (Khokhlov 1995; Niemeyer \& Woosley 1997)
and future investigations should include such a time dependence, which
is in agreement with the findings discussed above.

	Finally, nucleosynthesis can also give clues about the
deflagration-detonation transition (DDT). The most obvious consequence
of choosing different transition densities is the amount of $^{56}$Ni
produced in a SN Ia event.  H\"oflich and Khokhlov (1996) find from
light curve modeling and spectra that the typical $^{56}$Ni mass
should be in the range 0.5-0.7~M$_\odot$. This agrees with W7. Among
the DD models it would ask for a value somewhere between DD1 and DD2
(closer to DD2). Here 3, 2, and 1 stand for DDTs when densities ahead
of the flame decrease to 3.0, 2.2, and 1.7 \e7 \gmc~. DD1 is excluded
for other reasons. The amount of Si-Ca in comparison to Fe is too
large in DD1 models in order to compensate for the well-known
overproduction of Si-Ca in SNe II during galactic evolution.  Si/Fe
ratios in SN Ia models require specific Ia/(II+Ib) ratios in order to
obtain a solar mix combined with SN II contributions (see Table 5).
DD2 seems to be closest to the present observational constraints for
this ratio by Cappelaro et al. (1997).

	Small DDT densities favor larger amounts of matter that
experience incomplete Si burning.  Low-metallicity constraints require
an overproduction of Mn (and Cr) in SNe Ia. These elements are mostly
made as $^{55}$Co and $^{52}$Fe (decaying to Mn and Cr), which are
favorably produced in incomplete Si burning and would also require a
DDT between DD1 and DD2. (One should, however, realize that a fast
deflagration could possibly simulate this as well - see the Mn
overproduction in Figure 12 - and on the other hand that these numbers
would have to be rescaled or reinterpreted in multi-D calculations.)
Thus, combining all requirements on the DDT from total Ni yields,
Si/Fe and Ia/(II+Ib) ratios, as well as specific elements favored in
incomplete Si-burning, we would argue for a DDT density slightly below
2.2 \e7 \gmc~, i.e., results between DD1 and DD2. One should, however,
be careful with these constraints based on spherically symmetric
approximations of the burning front. Full three-dimensional
calculations could possibly produce the required ratio of matter from
incomplete Si burning and complete Si burning with alpha-rich
freeze-out in a different realization.

	More extended calculations that make use of the conclusions
presented here, including a time dependence of $v_{\rm def}/v_{\rm
sound}$, the best choice for the DDT density, and a detailed galactic
evolution model, replacing our comparisons of only the global yields,
will be the next step to undertake. That would also have to include
further tests of the metallicity of the exploding object, a topic
which has gained in importance with the cosmological interpretation of
high-redshift SNe Ia.  For this reason we repeat here in Figure
25a-25c some of the Figures 19-21 with a continuation to smaller mass
fractions and purely plotted as a function of velocity, in order to
magnify the behavior of the outer layers.  The differences in
velocities between CDD and WDD models are negligibly small in these
plots.


As mentioned before, the $^{54}$Fe (in the outer layers not affected
by electron captures but only by the neutron excess due to the initial
metallicity) ranging in velocities up to 15,000 - 19,000 km s$^{-1}$
in the models DD1-DD3, is a strong indicator of metallicity (see Fig.
14).  The minima in S, Ar, and Ca, if observed in spectra, with
positions between 16,000 and 21,000 km s$^{-1}$ in the DD1-DD3 models,
could give further clues on the deflagration-detonation
transition. And finally, any unburned intermediate mass elements at
higher velocities would give a clear indication of the metallicity of
the accreted matter. Our calculations and plots include only initial
compositions of $^{12}$C, $^{16}$O, and $^{22}$Ne. Solar abundances of
Ca, Ar, S, Si, or Fe would correspond to mass fractions of $1.2\times
10^{-4}$, $1.6 \times 10^{-4}$, $7.6\times 10^{-4}$, $1.3\times
10^{-3}$, or $2.3\times 10^{-3}$.

Thus, any future investigations of very early time spectra, leading to
abundance observations at high velocities, would provide strong
constraints on the SNe Ia mechanism and the relation to the
metallicity of the individual SNe Ia explosion.

\section*{Acknowledgments}

This work has been supported in part by the grant-in-Aid for
Scientific Research (05242102, 06233101) and COE research (07CE2002)
of the Ministry of Education, Science and Culture in Japan, a
fellowship of the Japan Society for the Promotion of Science for
Japanese Junior Scientists (6728), the Swiss Nationalfonds
(20-47252.96 and 2000-53798.98), the US National Science Foundation
(grant PHY94-07194), and the DOE (contract DE-AC05-96OR22464). Part of
the computations were carried out on a Fujitsu VPP-500 at the
Institute of Physical and Chemical Research (RIKEN) and the Institute
of Space and Astronautical Science (ISAS), and a Fujitsu VPP-300 at
the National Astronomical Observatory in Japan (NAO, Tokyo).  We want
to thank all participants of the NATO workshop on Thermonuclear
Supernovae in Aiguablava, where the idea to the present research
began.  Some of us (K.I., K.N., W.R.H., and F.K.T.) thank the ITP at
the University of California, Santa Barbara, for hospitality and
inspiration during the supernova program.  The paper was completed
during the SNe Ia workshop at the Aspen Center for Physics (June
1999).

\newpage

\epsfverbosetrue

\newpage

\newpage

\newpage

\vspace{2cm}

\newpage

\newpage

\begin{figure}[ht]
\epsfxsize=7cm
\epsfysize=6cm
\epsfbox[-900 86 -495 583]{xics15_central.epsi}
\end{figure}

\begin{figure}[hb]
\epsfxsize=7cm
\epsfysize=6cm
\epsfbox[-900 81 -495 614]{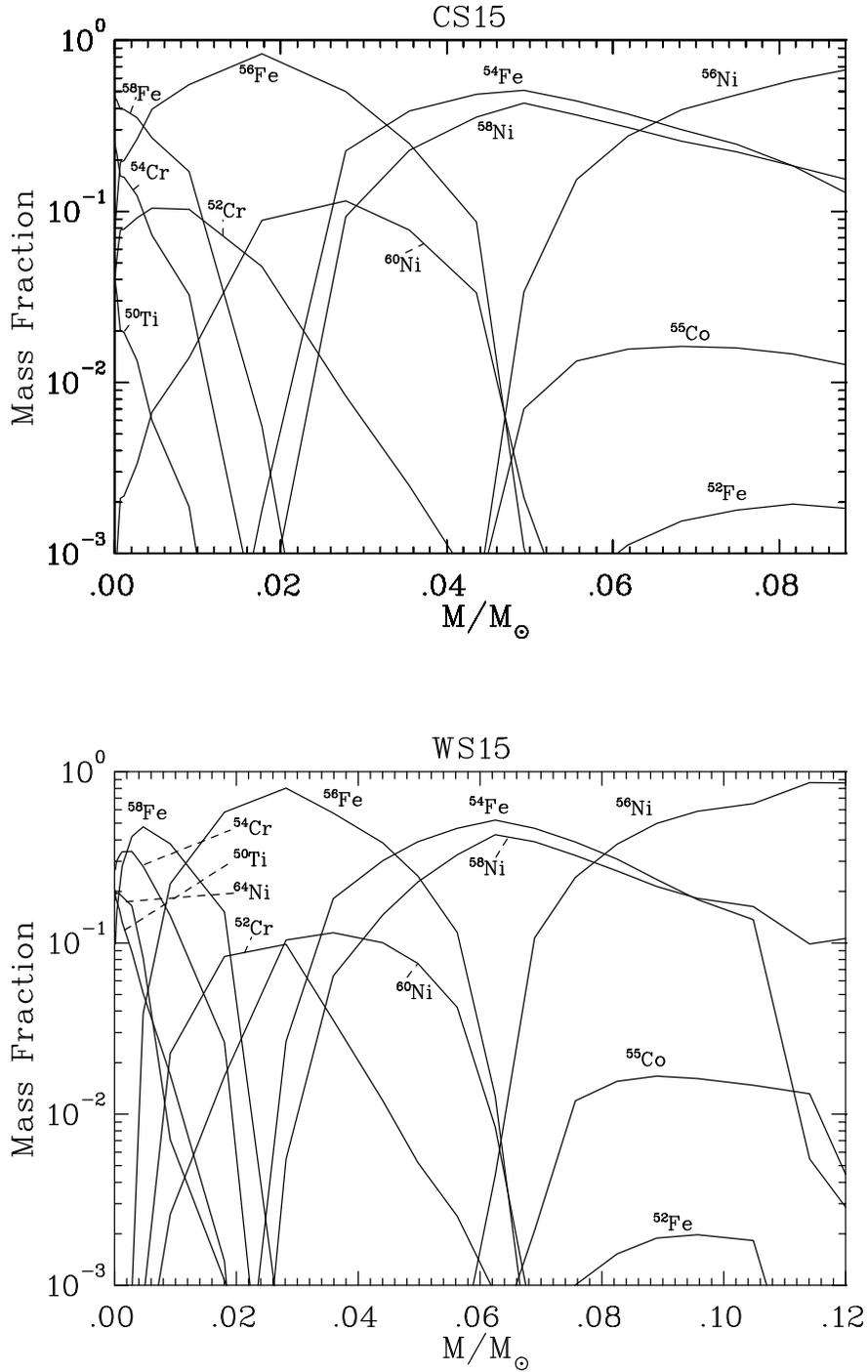}
\vspace{1cm}
\caption{ Abundance plots (mostly of Fe-group nuclei) for the cases
CS15 and WS15. Shown are the locations of the nuclei overproduced in
W7, $^{54}$Fe and $^{58}$Ni, corresponding to $Y_e$ values of
0.47-0.485. Due to the $Y_e$-gradients which are steeper than for W7,
the amount of matter in a given $Y_e$-range is reduced, but also
smaller central values are attained, giving rise to more neutron-rich
nuclei.  A $Y_e$ of about 0.46$\approx$26/56 (which was also attained
in W7) causes no problems and leads to a large abundance of stable
$^{56}$Fe (not from $^{56}$Ni decay). Values in the range of 0.44 to
0.46 result also in $^{50}$Ti, $^{54}$Cr, $^{58}$Fe, and $^{62,64}$Ni. 
$^{48}$Ca with Z/A$\approx$0.42 is only produced if $Y_e$ approaches
values smaller than 0.44 (see Hartmann et al. 1985). \label{xi15c}
}
\end{figure}

\newpage

\begin{figure}[ht]
\epsfxsize=7cm
\epsfysize=6cm
\epsfbox[-900 86 -495 583]{xics30_central.epsi}
\end{figure}

\begin{figure}[hb]
\epsfxsize=7cm
\epsfysize=6cm
\epsfbox[-900 81 -495 614]{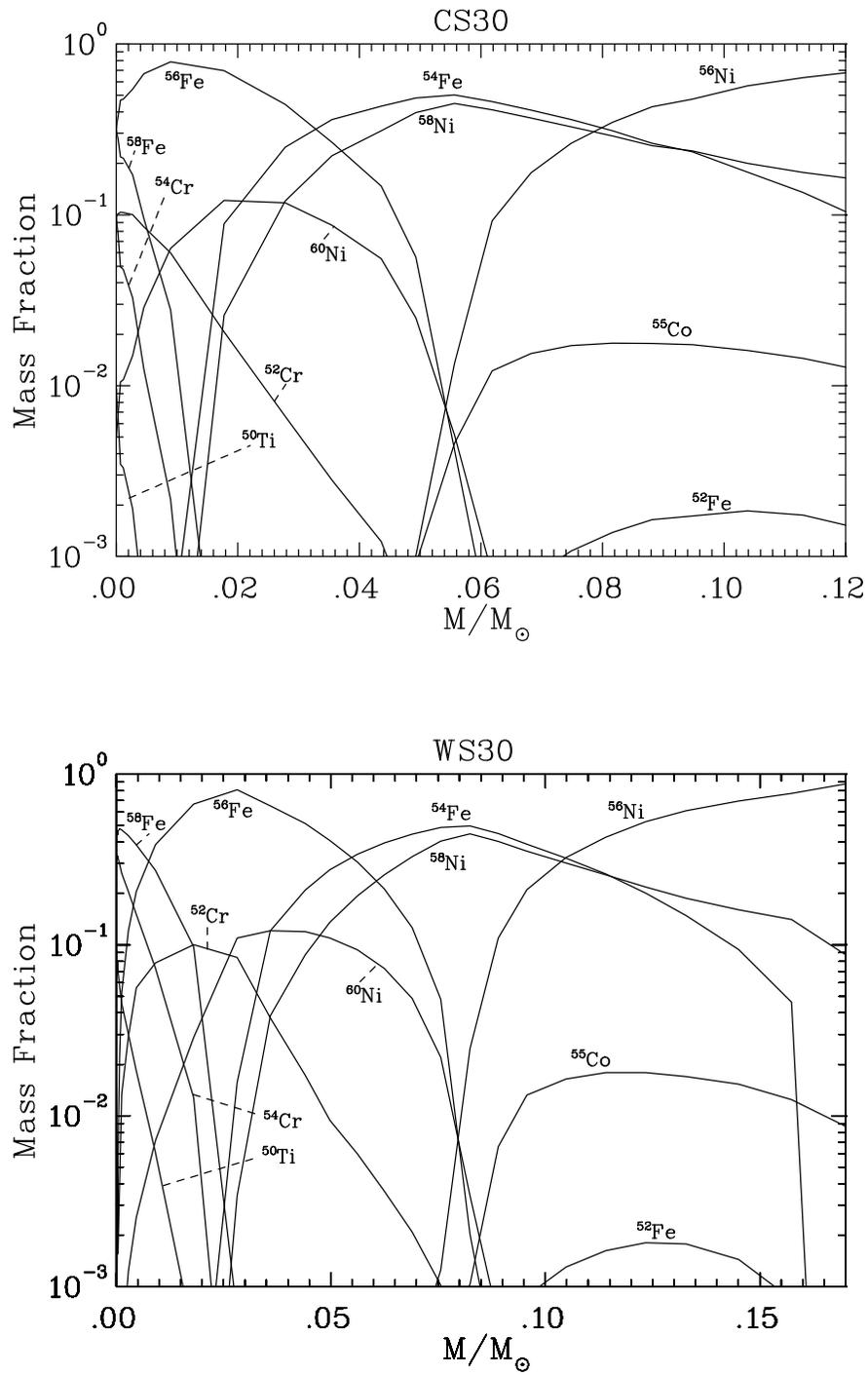}
\vspace{1cm}
\caption{ Abundance plots (mostly of Fe-group nuclei) for the cases
Same as \ref{xi15c} for the
cases CS30 and WS30. \label{xi30c}}
\end{figure}

\newpage

\begin{figure}[ht]
\epsfxsize=7cm
\epsfysize=6cm
\epsfbox[-900 86 -495 583]{xics50_central.epsi}
\end{figure}

\begin{figure}[hb]
\epsfxsize=7cm
\epsfysize=6cm
\epsfbox[-900 81 -495 614]{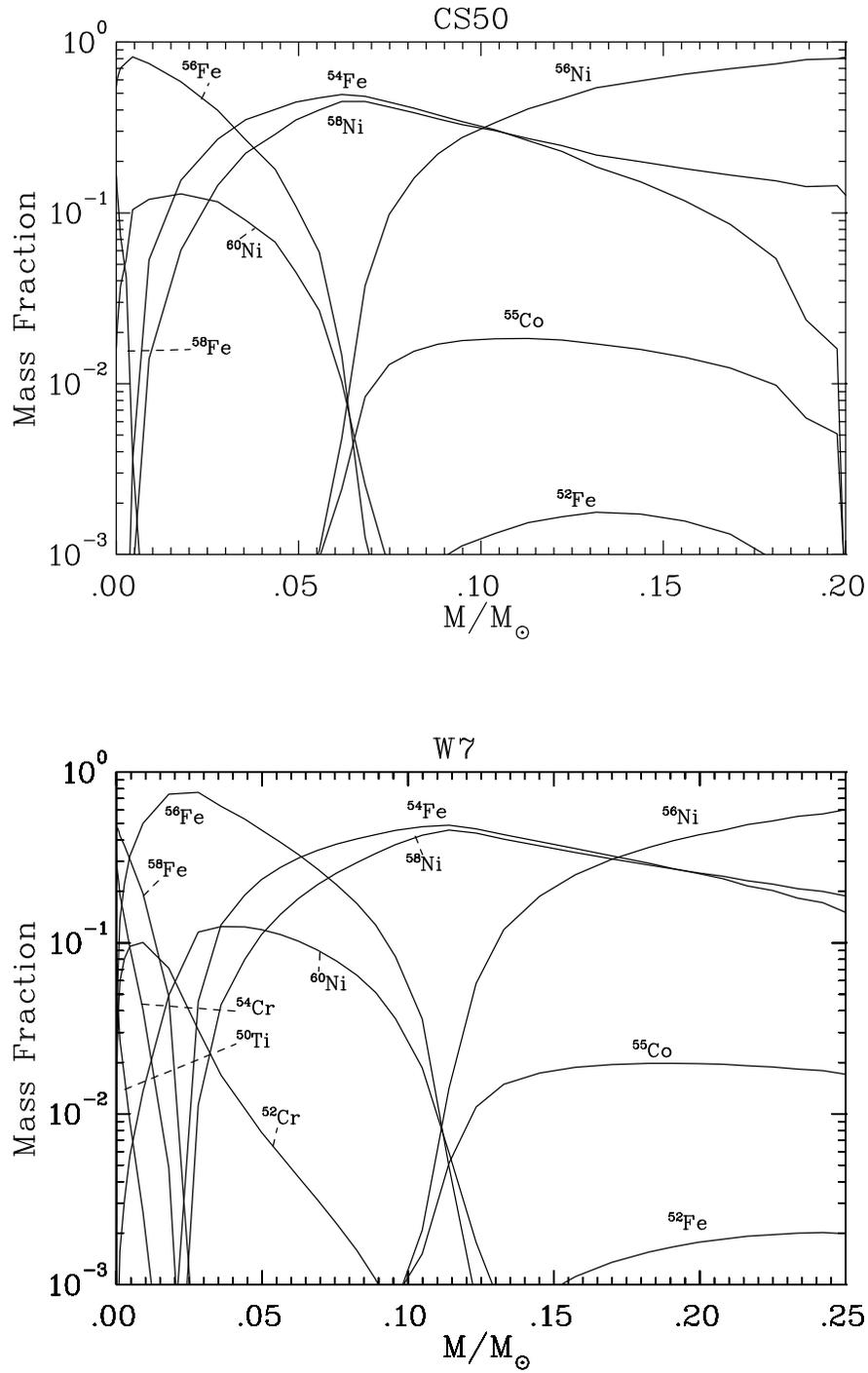}
\vspace{1cm}
\caption{ 
Same as \ref{xi15c} for the cases CS50 and W7. 
One sees that the low $Y_e$ region becomes more extended with increasing
burning front velocities, while the central values increase somewhat.
$^{55}$Co and $^{52}$Fe are not plotted here for W7 but are are
shown in Figure \ref{w7full}.
\label{xiw7c}} 
\end{figure}

\newpage

\begin{figure}[ht]
\epsfxsize=7cm
\epsfysize=6cm
\epsfbox[-900 86 -495 583]{xiwsl_central.epsi}
\end{figure}

\begin{figure}[hb]
\epsfxsize=7cm
\epsfysize=6cm
\epsfbox[-900 81 -495 614]{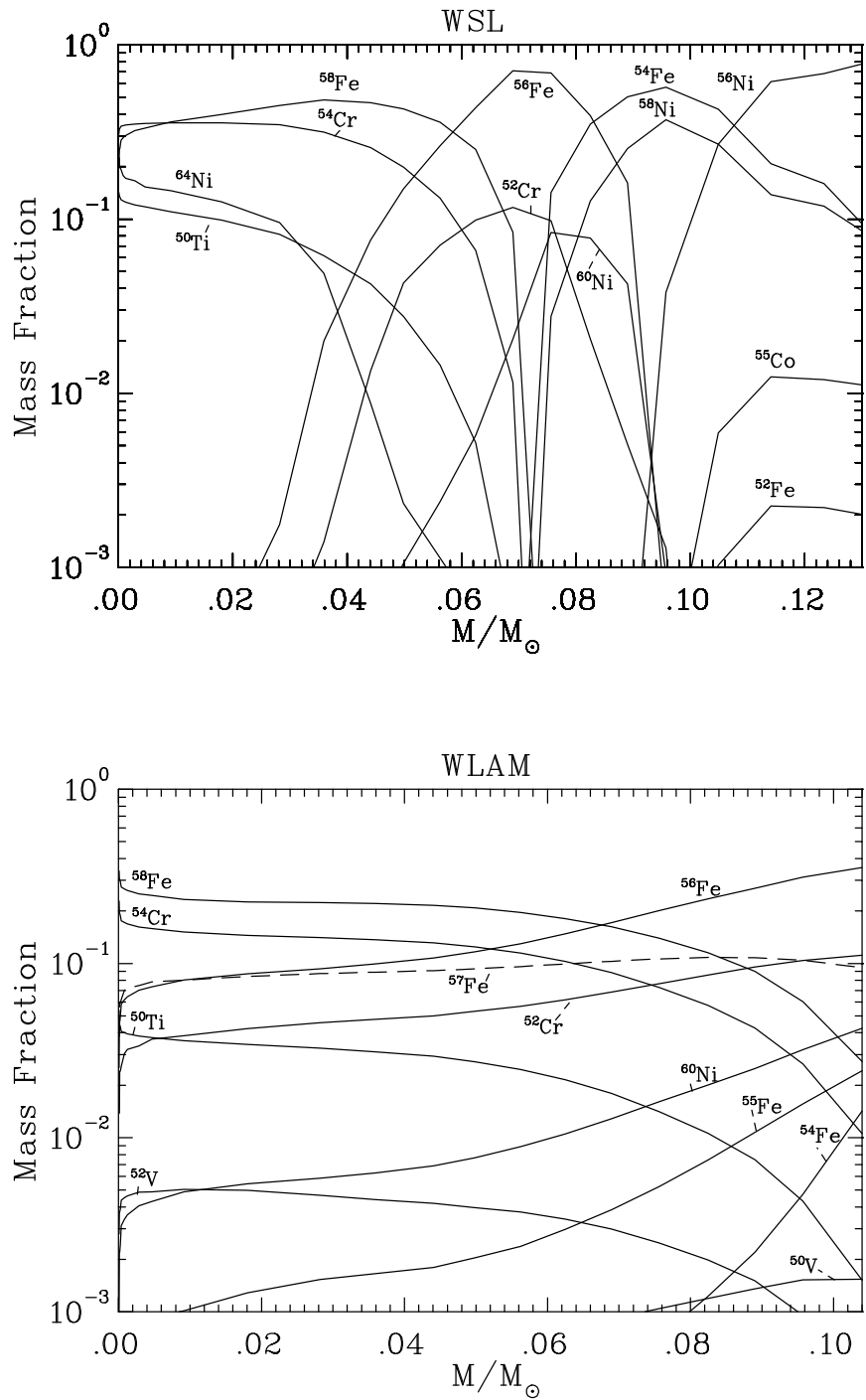}
\vspace{1cm}
\caption{ 
Same as \ref{xi15c} for the
cases WSL and WLAM. WLAM has an almost constant $Y_e$(\mr) with values
close to those of WSL at 0.05 M$_\odot$ (see also Figure
\ref{yemr}). \label{xiwslc}}
\end{figure}

\newpage

\begin{figure}[ht]
\epsfxsize=7cm
\epsfysize=7cm
\epsfbox[-950 -100 -472 520]{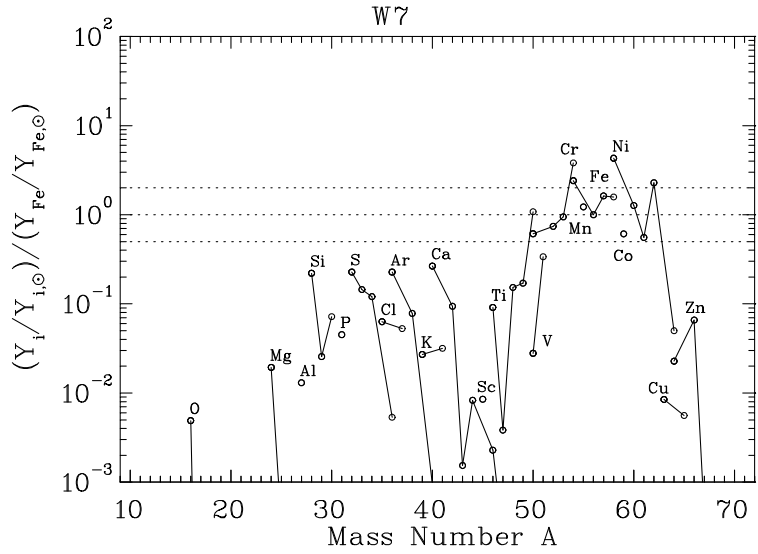}
\vspace{-1.5cm}
\caption{ 
Ratio of abundances to solar predicted in model W7 (this
is a recalculation of the 1986 model [Thielemann, Nomoto \& Yokoi
1986] with presently available updated reaction rates and a screened
NSE treatment for temperatures beyond $6\times 10^9$K, as described in
Hix \& Thielemann 1996). Isotopes of one element are connected by
lines. The ordinate is normalized to $^{56}$Fe. Intermediate mass
elements exist, but are underproduced by a factor of 2-3 in comparison
to Fe-group elements. Among the Fe-group, $^{54}$Cr and $^{58}$Ni are
overproduced by a factor of 4, which exceeds the permitted factor of
\(\sim 3\).\label{w7sol}}
\end{figure}

\begin{figure}[hb]
\epsfxsize=7cm
\epsfysize=7cm
\epsfbox[-950 -100 -472 520]{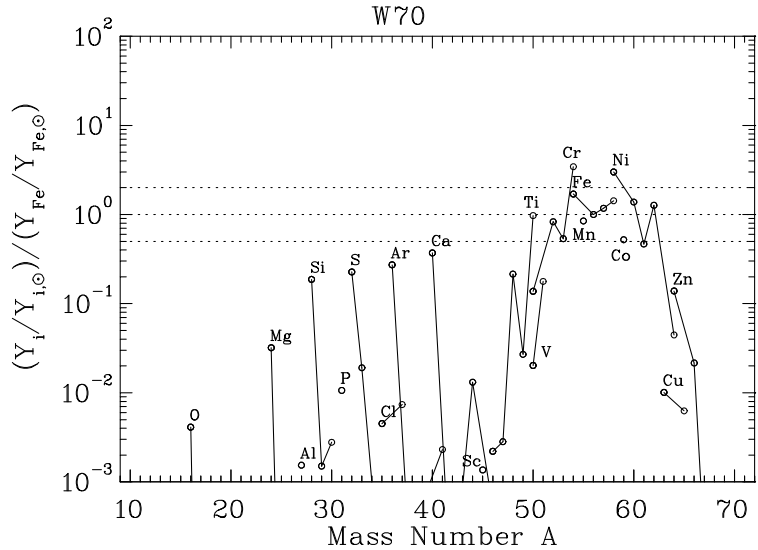}
\vspace{-2cm}
\caption{Same as Figure \ref{w7sol}, i.e. a deflagration model
treated exactly like W7, but with an initial composition corresponding
to zero metallicity (50\% mass fractions of $^{12}$C and $^{16}$O
without $^{22}$Ne admixture, i.e. $Y_e$=0.5). The production of
neutron-rich species $^{54}$Fe, $^{58}$Ni, and $^{62}$Ni are smaller
than the original W7 model. Noticeable is also the almost complete
absence of intermediate odd-Z elements which are produced in explosive
oxygen burning for a $Y_e$ smaller than 0.5. The even-Z intermediate
mass elements are also effected. They are not set off by a constant
factor in comparison to $^{56}$Fe, as seen in Figure \ref{w7sol} for
W7, but increase with mass. The increasing deviation between $N=Z$ and
a line in the nuclear chart with $Y_e$$<$0.5, with increasing nuclear
mass, does not occur. \label{w70sol}}
\end{figure}

\newpage

\begin{figure}[hb]
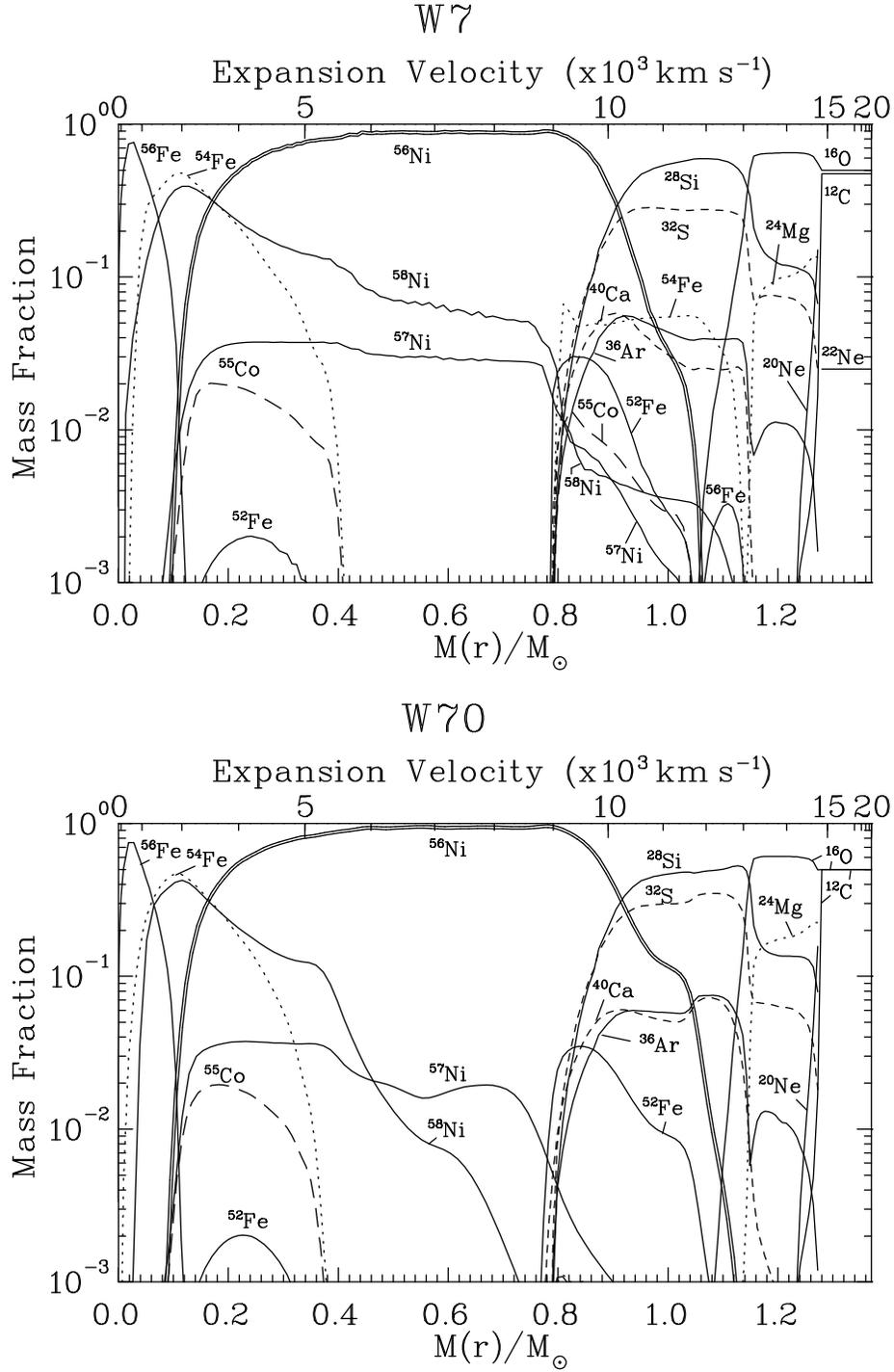

\epsfxsize=7cm
\epsfysize=7cm
\epsfbox[-900 120 -512 570]{xiw7_full.epsi}
\vspace{1.5cm}
\epsfxsize=7cm
\epsfysize=7cm
\epsfbox[-900 120 -512 570]{xiw70_full.epsi}
\vspace{1.5cm}
\caption{Composition of W7 and W70. The major changes are given in the mass
zones outside the central 0.3\ms, where $Y_e$ is enherited from the initial
metallicity in form of $^{22}$Ne and not due to electron captures.
This has effects on the Fe-group composition in the alpha-rich freeze-out
region (on $^{58,57}$Ni and $^{52}$Fe, the latter two decaying to 
$^{57}$Fe and $^{52}$Cr) and the incomplete Si-burning region
($^{58,57}$Ni, $^{55}$Co, and $^{54,52}$Fe, the unstable species decaying
to $^{57}$Fe, $^{55}$Mn, and $^{52}$Cr). 
$^{58,57}$Ni, $^{55}$Co, and $^{54}$Fe decrease below the plot limits
in the incomplete burning region while $^{52}$Fe (N=Z) increases.
\label{w7full}}
\end{figure}

\newpage

\begin{figure}[hb]
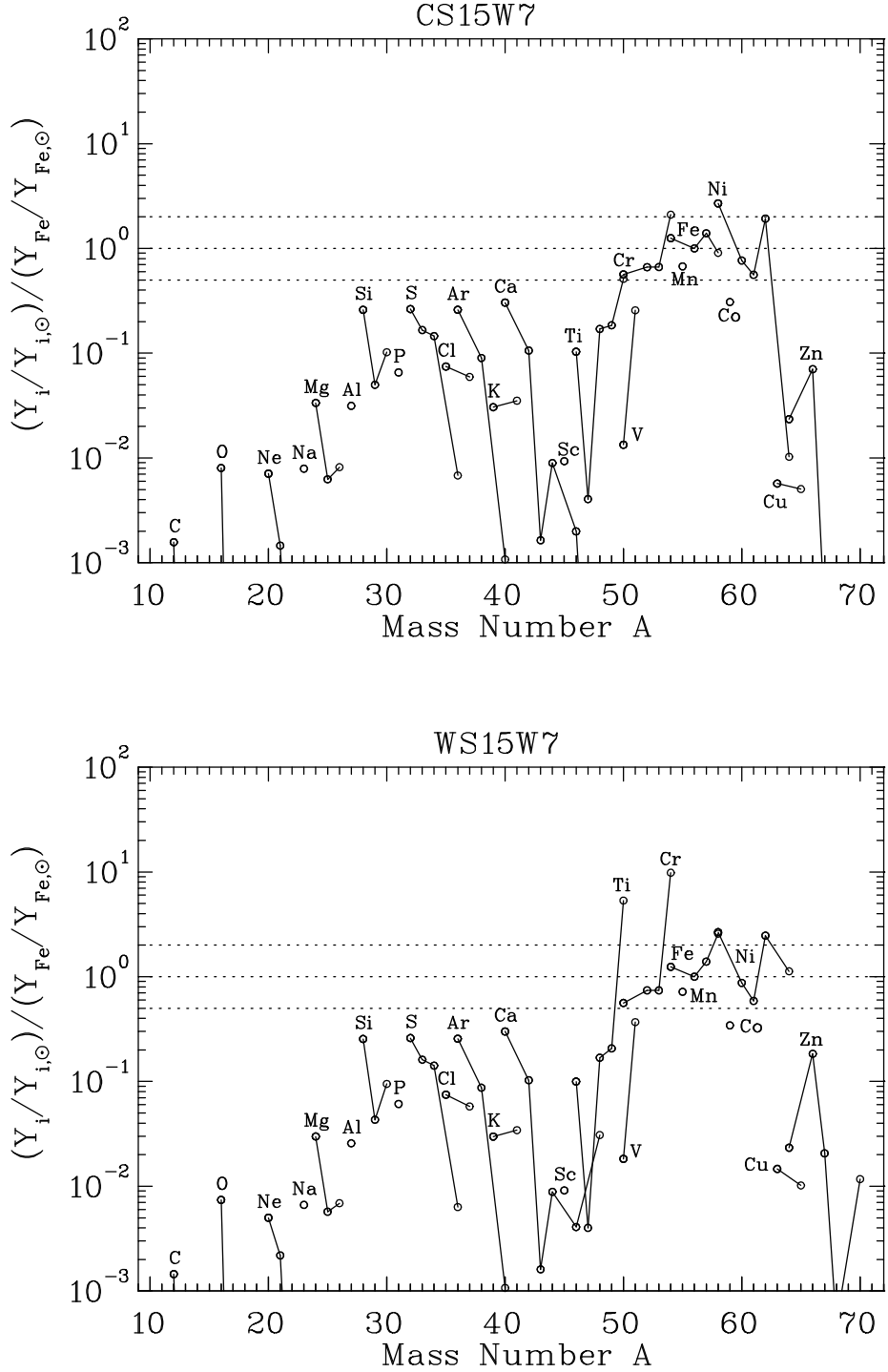

\epsfxsize=7cm
\epsfysize=7cm
\epsfbox[-900 120 -512 570]{ysolcs15w7.epsi}
\vspace{2cm}
\epsfxsize=7cm
\epsfysize=7cm
\epsfbox[-900 120 -512 570]{ysolws15w7.epsi}
\vspace{2cm}
\caption{Comparison to solar for SN Ia compositions which are made up of
the slow deflagrations CS15 and WS15 in the central layers and a W7
composition in the outer layers. This is mainly a test for the
Fe-group composition, especially $^{50}$Ti, $^{54}$Cr, $^{58}$Fe,
and $^{64}$Ni. \label{s15solw7}}
\end{figure}

\newpage

\begin{figure}[hb]
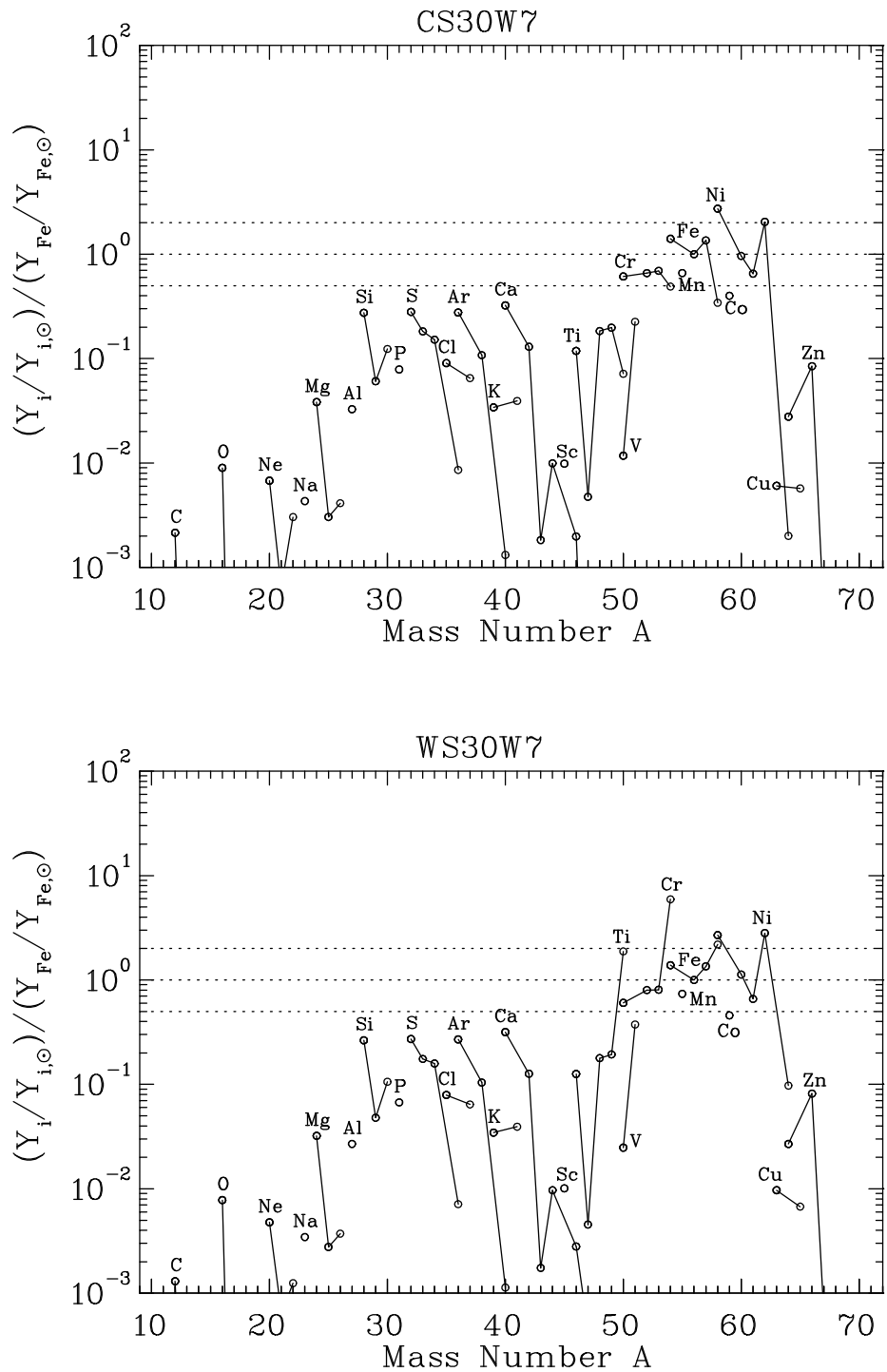

\epsfxsize=7cm
\epsfysize=7cm
\epsfbox[-900 120 -512 570]{ysolcs30w7.epsi}
\vspace{2cm}
\epsfxsize=7cm
\epsfysize=7cm
\epsfbox[-900 120 -512 570]{ysolws30w7.epsi}
\vspace{2cm}
\caption{Comparison to solar for a SN Ia composition made up from
CS30 and WS30 in the central layers and W7 in the outer layers. 
The neutron-rich species decline in comparison with CS15 and WS15.
\label{s30solw7}}
\end{figure}

\newpage

\begin{figure}[hb]
\epsfxsize=7cm
\epsfysize=7cm
\vspace{3cm}
\epsfbox[-900 320 -512 770]{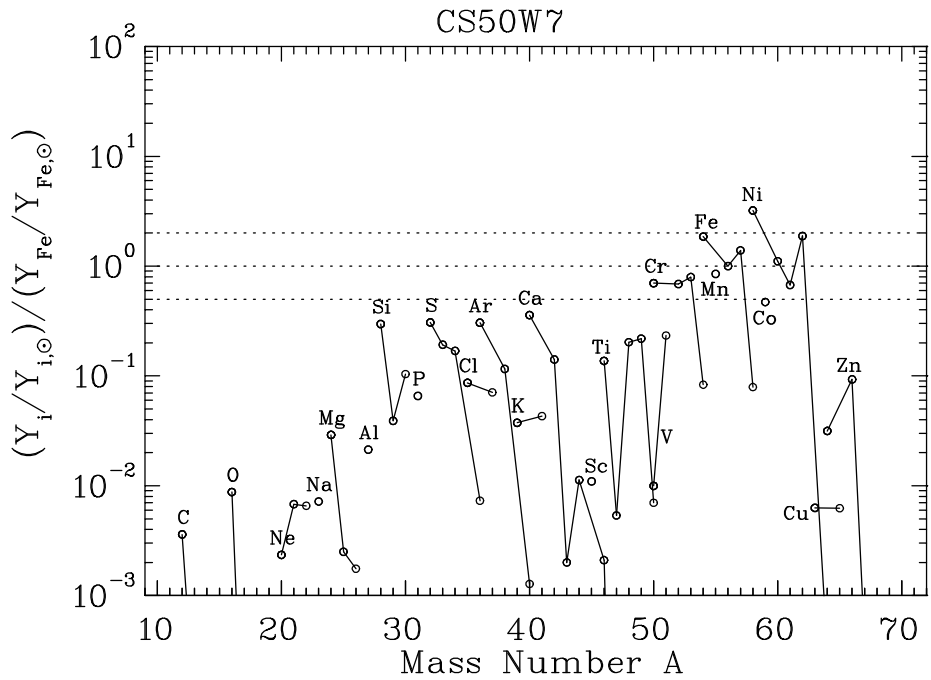}
\vspace{6cm}
\caption{Comparison to solar for CS50 with a W7 composition in
the outer layers. A further decrease of (very) neutron-rich species is visible,
but $^{54}$Fe and $^{58}$Ni, originating from intermediate $Y_e$ regions,
increase to excessive values due to the flatter $Y_e$-gradient of faster
deflagrations (see Figure \ref{yemr}).
\label{s50solw7}}
\end{figure}

\newpage

\begin{figure}[hb]
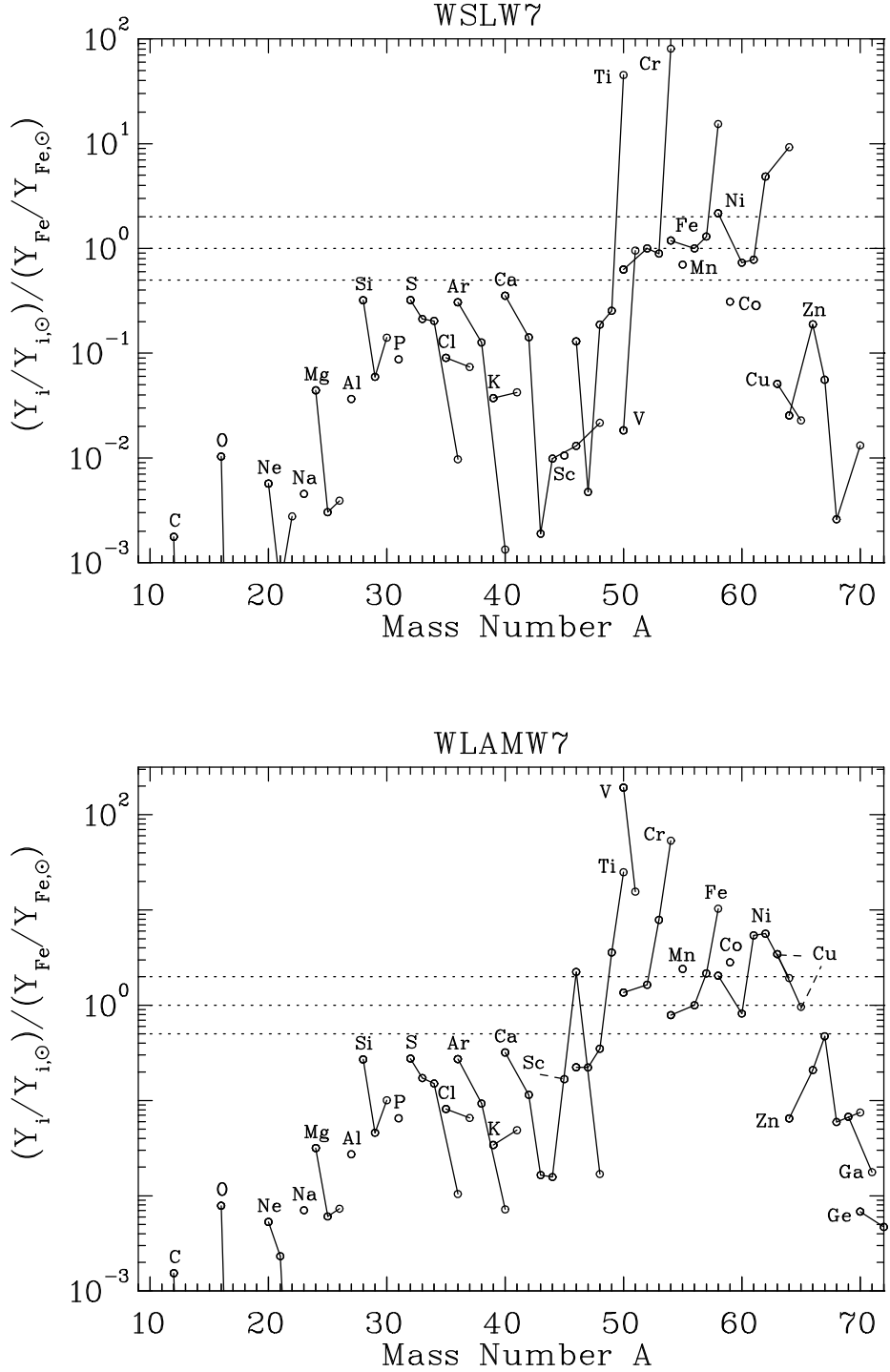

\epsfxsize=7cm
\epsfysize=7cm
\epsfbox[-900 120 -512 570]{ysolwslw7.epsi}
\vspace{2cm}
\epsfxsize=7cm
\epsfysize=7cm
\epsfbox[-900 120 -512 570]{ysolwlamw7.epsi}
\vspace{2cm}
\caption{Comparison to solar for WSL and WLAM with a W7 composition in
the outer layers. Excessive ratios for neutron-rich species are observed
due to extended low $Y_e$-regions out to 0.06 M$_\odot$ (see Figure
\ref{yemr}).
\label{slsolw7}}
\end{figure}

\newpage

\begin{figure}[hb]
\epsfxsize=7cm
\epsfysize=7cm
\epsfbox[-900 120 -512 570]{xiws15dd1_full.epsi}
\vspace{2cm}
\epsfxsize=7cm
\epsfysize=7cm
\epsfbox[-900 120 -512 570]{xiws15dd2_full.epsi}
\vspace{2cm}
\caption{Composition of WS15DD1 and WS15DD2 against the expansion
velocity and $M(r)$. \label{wdd12}}
\end{figure}

\newpage

\begin{figure}[hb]
\epsfxsize=7cm
\epsfysize=7cm
\vspace{3cm}
\epsfbox[-900 320 -512 770]{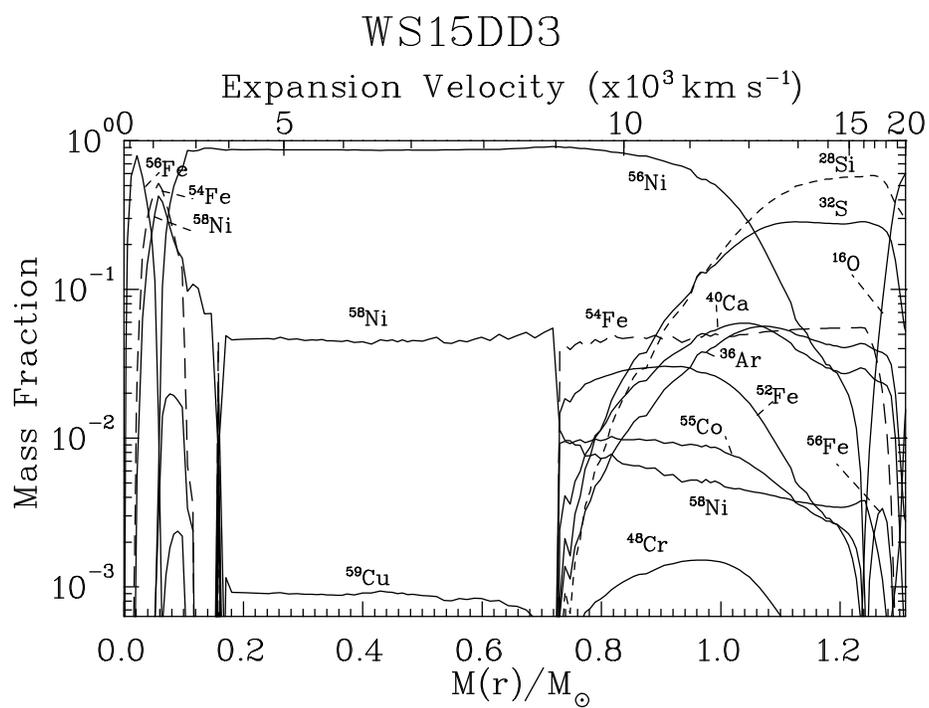}
\vspace{6cm}
\caption{Composition of WS15DD3.
In the series DD1-DD3 we
see a decrease in the total amount of intermediate mass elements (Si-Ca), an
increase in $^{56}$Ni, and a change of the ratio between matter experiencing
an alpha-rich freeze-out (indicated by the $^{58}$Ni-plateau) and
incomplete Si-burning ($^{54}$Fe-plateau). 
 \label{wdd13}}

\end{figure}

\newpage

\begin{figure}[hb]
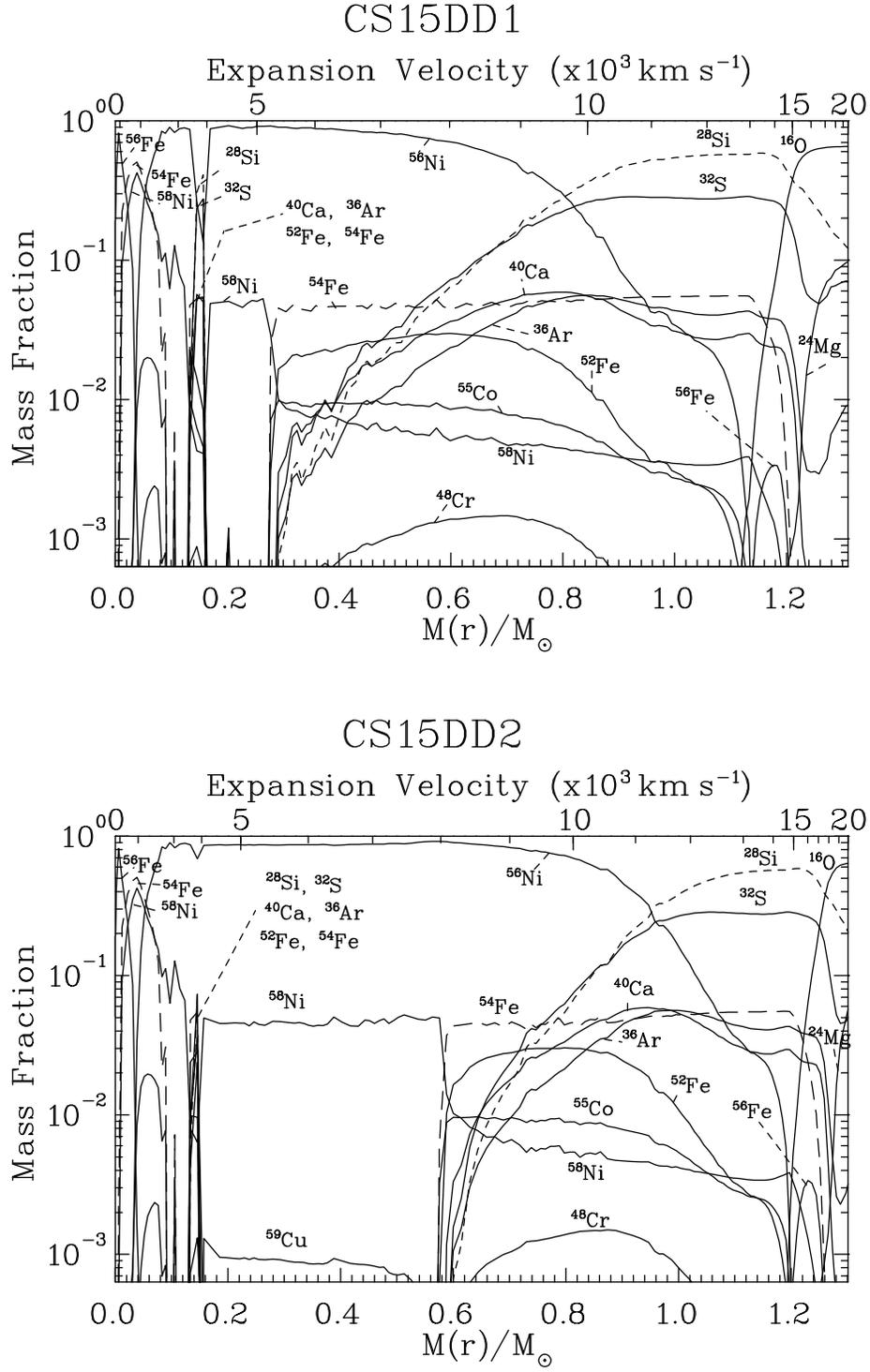

\epsfxsize=7cm
\epsfysize=7cm
\epsfbox[-900 120 -512 570]{xics15dd1_full.epsi}
\vspace{2cm}
\epsfxsize=7cm
\epsfysize=7cm
\epsfbox[-900 120 -512 570]{xics15dd2_full.epsi}
\vspace{2cm}
\caption{Composition of CS15DD1 and CS15DD2. 
The outer layers affected by the detonation behave almost identical to the
WSDD series.
 \label{cdd12}}
\end{figure}

\newpage

\begin{figure}[hb]
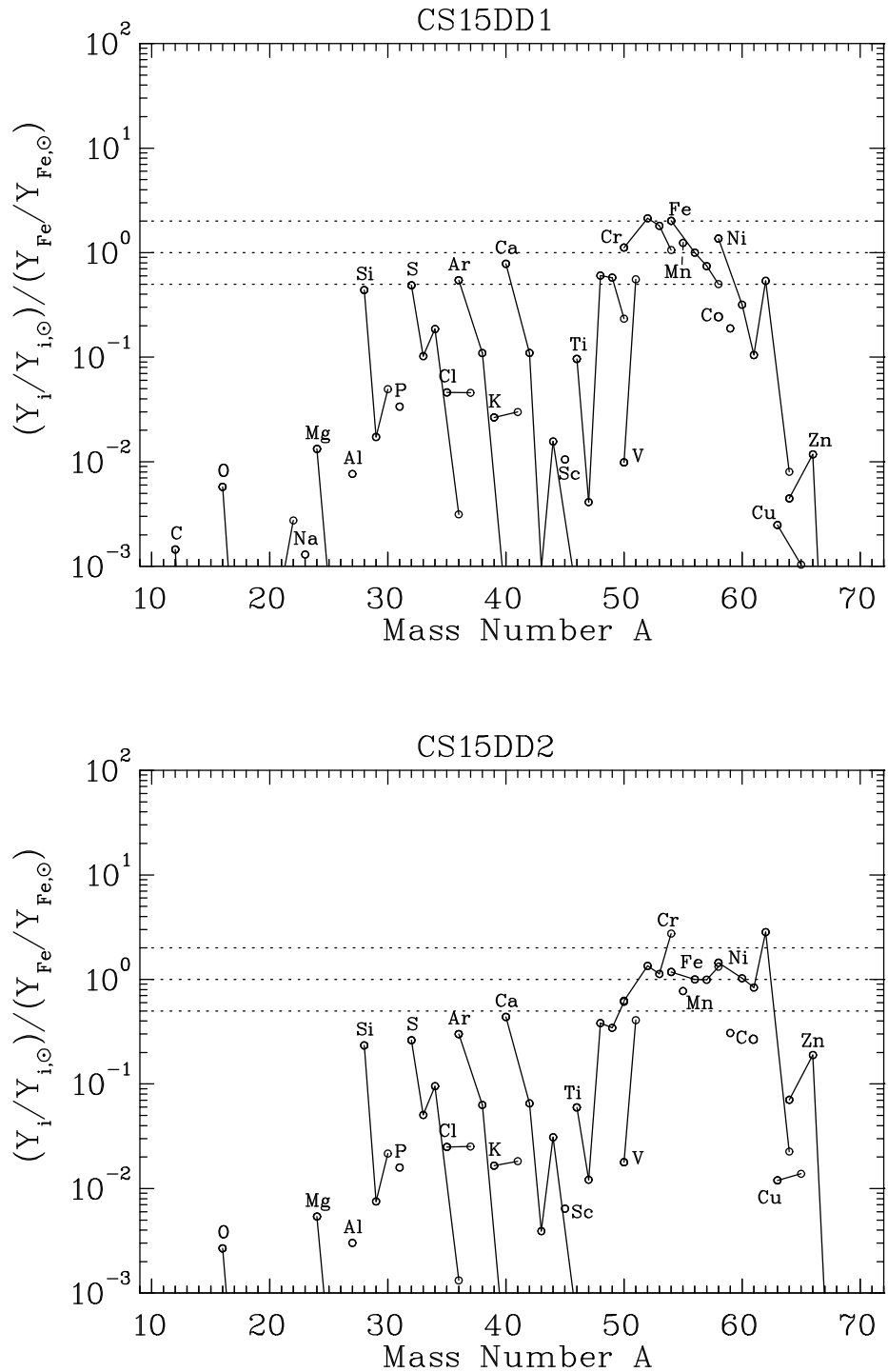

\epsfxsize=7cm
\epsfysize=7cm
\epsfbox[-900 120 -512 570]{ysolcs15dd1.epsi}
\vspace{2cm}
\epsfxsize=7cm
\epsfysize=7cm
\epsfbox[-900 120 -512 570]{ysolcs15dd2.epsi}
\vspace{2cm}
\caption{Comparison to solar for CS15DD1 and CS15DD2. A decrease
of the Si-Ca/Fe ratio is observable, a decrease of e.g. $^{55}$Mn and
$^{52}$Cr (dominated by incomplete Si-burning), and an increase of
$^{62}$Ni (decaying from $^{62}$Zn) and $^{59}$Co (alpha-rich
freeze-out).  
\label{csdsol}}

\end{figure}

\newpage

\begin{figure}[hb]
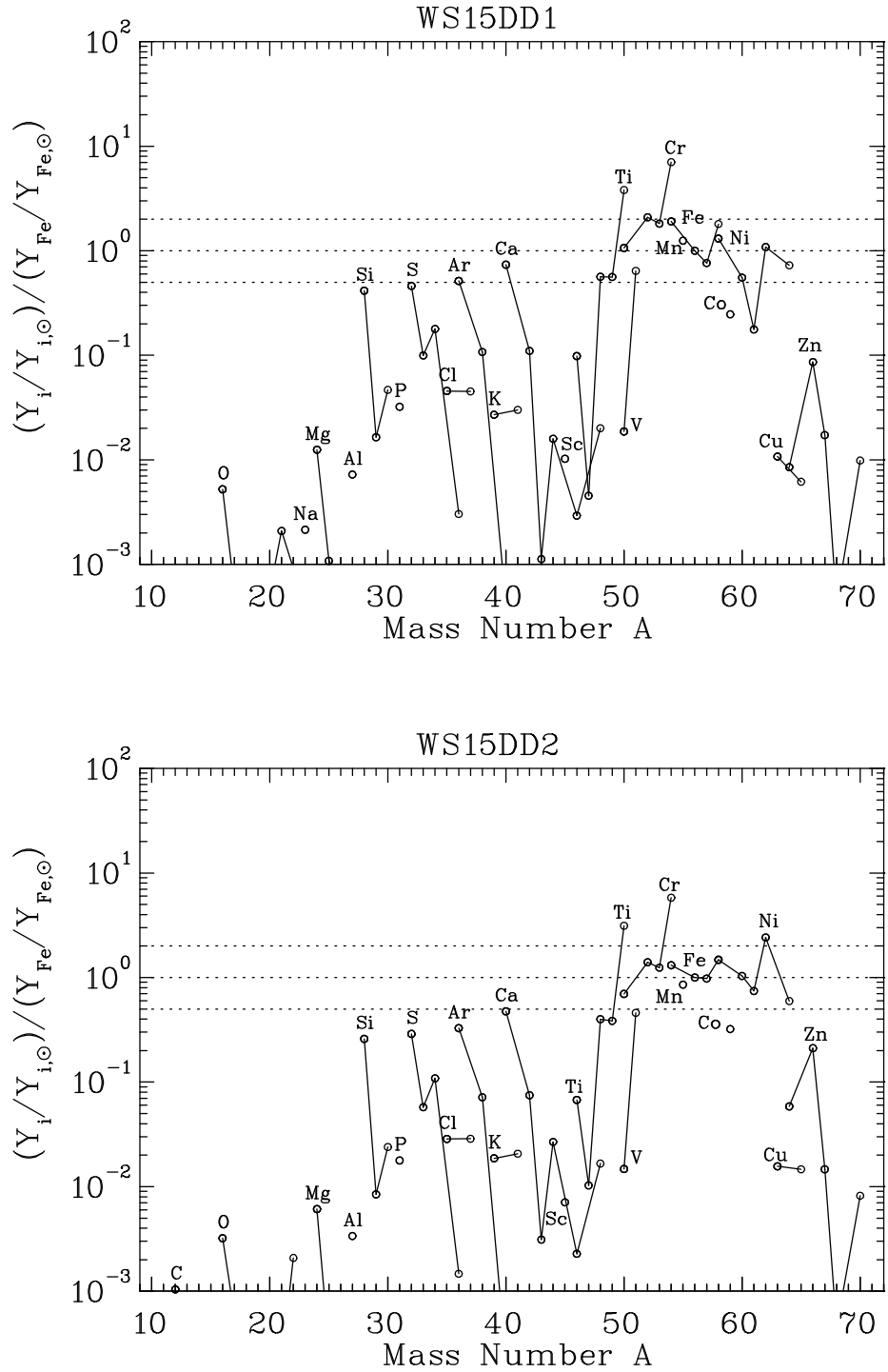

\epsfxsize=7cm
\epsfysize=7cm
\epsfbox[-900 120 -512 570]{ysolws15dd1.epsi}
\vspace{2cm}
\epsfxsize=7cm
\epsfysize=7cm
\epsfbox[-900 120 -512 570]{ysolws15dd2.epsi}
\vspace{2cm}
\caption{Comparison to solar for WS15DD1 and WS15DD2. \label{wsdsol}}

\end{figure}

\newpage

\begin{figure}[hb]
\epsfxsize=7cm
\epsfysize=7cm
\vspace{3cm}
\epsfbox[-900 320 -512 770]{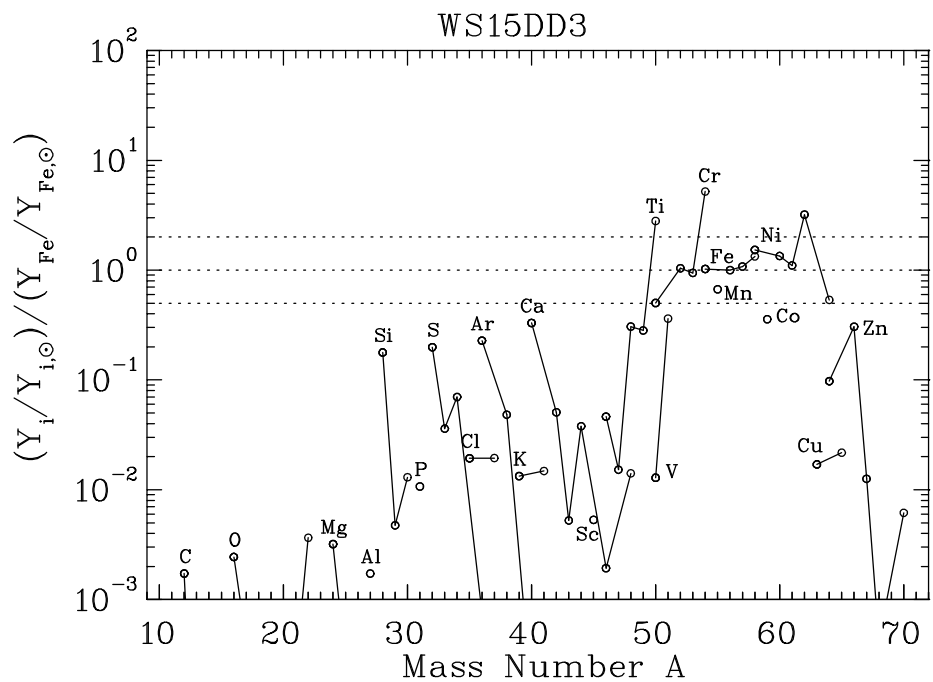}
\vspace{6cm}
\caption{Comparison to solar for WS15DD3. In this series DD1-DD3
we see similar changes as shown in Figure \ref{csdsol}. Differences in
the neutron-rich species $^{50}$Ti, $^{54}$Cr, $^{58}$Fe, and
$^{62}$Ni are due to the differences between the W and C series in the
central slow deflagration layers (higher ignition densities).
\label{wsd3sol}}
\end{figure}

\newpage

\begin{figure}[hb]
\epsfxsize=7cm
\epsfysize=5cm
\vspace*{-3cm}
\epsfbox[-1000 120 -502 770]{xivelcdd1.ps}
\epsfxsize=7cm
\epsfysize=5cm
\vspace{-2cm}
\epsfbox[-1000 120 -502 770]{xivelcdd2.ps}
\epsfxsize=7cm
\epsfysize=5cm
\vspace{-2cm}
\epsfbox[-1000 120 -502 770]{xivelcdd3.ps}
\vspace{0.5cm}
\caption{Mass fractions in the models CS15DD1-DD3 as a function of expansion
velocity, which reveals more easily the outer layers affected by the
deflagration-detonation transition and the quenching of nuclear
burning up to unburned matter in the very outer layers. Intermediate
mass elements beyond Mg are essentially unburned for velocities larger
than 20000-30000 km s$^{-1}$ and would be found in their solar values
scaled with the appropriate metallcity. The initial composition of the
white dwarf for the present calculation consists only of $^{12}$C,
$^{16}$O, and $^{22}$Ne (according to the CNO metallicity of the
accreted matter).
\label{xivel}}

\end{figure}

\newpage

\begin{table}
\centerline{Table 1: Masses of $^{56}$Ni in SN Ia Models Investigated}
\vskip 0.2cm
\label{tabdd}
\begin{center}
\begin{tabular}{crrrr}
\hline
Model & Mass & Energy & $^{56}$Ni Mass \\
\hline
W7   & 1.38 \ms\ & 1.3\afoe & 0.59 \ms\ \\
W70  & 1.38 \ms\ & 1.3\afoe & 0.64 \ms\ \\
WS15DD1 & 1.38 \ms\ & 1.33\afoe & 0.56 \ms\ \\
WS15DD2 & 1.38 \ms\ & 1.40\afoe & 0.69 \ms\ \\
WS15DD3 & 1.38 \ms\ & 1.43\afoe & 0.77 \ms\ \\
CS15DD1 & 1.38 \ms\ & 1.34\afoe & 0.56 \ms\ \\
CS15DD2 & 1.38 \ms\ & 1.44\afoe & 0.74 \ms\ \\
\hline
\end{tabular}
\end{center}
\end{table}

\begin{table}
\centerline{Table 2: Nucleosynthesis of Neutron-rich Species ($X_i/X(^{56}$Fe)/Solar Ratio)}
\vskip 0.2cm
\label{tabsd}
\begin{center}
\begin{tabular}{crrrrrrrrrr}
\hline
Model & $\rho_9$ & $v_{\rm def}$/$v_{\rm s}$ & $Y_{\rm e,c}$
& $^{50}$Ti & $^{54}$Cr & $^{58}$Fe & $^{64}$Ni & $^{62}$Ni & 
$^{54}$Fe & $^{58}$Ni \\
\noalign{\hrule}
WSL     & 2.12 & $\sim$0.01 & 0.442 & 39     & 74    & 15    & 7      & 5   & 1.2 & 2.1 \\
WLAM   & 2.12 & $\sim$0.01 & 0.447 & 22     & 47    &  9    & 1.6    & 5   & 1.8 & 3.8 \\
WS15W7  & 2.12 & 0.015      & 0.440 &  4     &  7    &  1.7  & 0.7    & 2.1 & 1.3 & 2.7 \\
WS15DD1 & 2.12 & 0.015      & 0.440 &  4     &  7    &  1.8  & 0.7    & 1.1 & 1.9 & 1.3 \\    
WS15DD2 & 2.12 & 0.015      & 0.440 &  3.1   &  6    &  1.5  & 0.6    & 2.4 & 1.3 & 1.5 \\
WS15DD3 & 2.12 & 0.015      & 0.440 &  2.8   &  5    &  1.3  & 0.5    & 3.2 & 1.0 & 1.5 \\
WS30W7  & 2.12 & 0.03       & 0.445 &  1.3   &  4    &  1.5  & 0.1    & 2.5 & 1.3 & 2.6 \\
CS15W7  & 1.37 & 0.015      & 0.449 &  0.5   &  2    &  0.9  & 0.01   & 1.9 & 1.3 & 2.7 \\
CS15DD1 & 1.37 & 0.015      & 0.449 &  0.2   &  1.1  &  0.5  & 0.01   & 0.5 & 2.0 & 1.4 \\
CS15DD2 & 1.37 & 0.015      & 0.449 &  0.6   &  2.7  &  1.3  & 0.01   & 2.8 & 1.2 & 1.4 \\
CS30W7  & 1.37 & 0.03       & 0.455 &  0.07  &  0.5  &  0.3  & 0.002  & 2.0 & 1.4 & 2.7 \\
CS50W7  & 1.37 & 0.05       & 0.459 &  0.003 &  0.04 &  0.04 & 0.0004 & 1.8 & 1.8 & 3.2 \\
W7      & 2.12 & up to 0.3  & 0.446 &  1.1   &  3.8  &  1.6  & 0.05   & 2.3 & 2.4 & 4.3 \\
W70     & 2.12 & up to 0.3  & 0.446 &  0.97  &  3.4  &  1.4  & 0.04   & 1.3 & 1.7 & 3.0 \\
\hline
\end{tabular}
\end{center}
\end{table}

\begin{table}
\centerline{Table 3: Nucleosynthesis products of SN II and Ia models
}
\vskip 0.2cm
\label{tabmas}
\begin{center}
\begin{tabular}{lcccccccc}
\hline \hline
& \multicolumn{7}{c}{Synthesized mass (M$_\odot$)} \\
\cline{2-9} 
& \multicolumn{1}{c}{Type II} & \multicolumn{7}{c}{Type Ia} \\
Species & 10$-$50M$_\odot$ & W7 & W70 & WDD1 & WDD2 & WDD3 & CDD1 & CDD2 \\
\hline 
$^{12}$C  & 7.93E-02 & 4.83E-02 & 5.08E-02 & 5.42E-03 & 8.99E-03 & 1.66E-02 & 9.93E-03 & 5.08E-03 \\
$^{13}$C  & 3.80E-09 & 1.40E-06 & 1.56E-09 & 5.06E-07 & 3.30E-07 & 3.17E-07 & 8.46E-07 & 4.16E-07 \\
$^{14}$N  & 1.56E-03 & 1.16E-06 & 3.31E-08 & 2.84E-04 & 2.69E-04 & 1.82E-04 & 9.06E-05 & 9.03E-05 \\
$^{15}$N  & 1.66E-08 & 1.32E-09 & 4.13E-07 & 9.99E-07 & 5.32E-07 & 1.21E-07 & 2.53E-07 & 2.47E-07 \\
$^{16}$O  & 1.80     & 1.43E-01 & 1.33E-01 & 8.82E-02 & 6.58E-02 & 5.58E-02 & 9.34E-02 & 5.83E-02 \\
$^{17}$O  & 9.88E-08 & 3.54E-08 & 3.33E-10 & 3.77E-06 & 4.58E-06 & 3.60E-06 & 9.55E-07 & 1.01E-06 \\
$^{18}$O  & 4.61E-03 & 8.25E-10 & 2.69E-10 & 6.88E-07 & 6.35E-07 & 2.39E-07 & 2.08E-07 & 1.92E-07 \\
$^{19}$F  & 1.16E-09 & 5.67E-10 & 1.37E-10 & 1.70E-09 & 4.50E-10 & 2.30E-10 & 5.83E-10 & 4.24E-10 \\
$^{20}$Ne & 2.12E-01 & 2.02E-03 & 2.29E-03 & 1.29E-03 & 6.22E-04 & 4.55E-04 & 1.16E-03 & 6.05E-04 \\
$^{21}$Ne & 1.08E-03 & 8.46E-06 & 2.81E-08 & 1.16E-05 & 1.39E-06 & 1.72E-06 & 3.63E-06 & 1.99E-06 \\
$^{22}$Ne & 1.83E-02 & 2.49E-03 & 2.15E-08 & 1.51E-04 & 4.21E-04 & 8.25E-04 & 4.41E-04 & 2.11E-04 \\
$^{23}$Na & 6.51E-03 & 6.32E-05 & 1.41E-05 & 8.77E-05 & 2.61E-05 & 3.01E-05 & 5.10E-05 & 3.50E-05 \\
$^{24}$Mg & 8.83E-02 & 8.50E-03 & 1.58E-02 & 7.55E-03 & 4.47E-03 & 2.62E-03 & 7.72E-03 & 4.20E-03 \\
$^{25}$Mg & 1.44E-02 & 4.05E-05 & 1.64E-07 & 8.23E-05 & 2.66E-05 & 2.68E-05 & 4.85E-05 & 3.25E-05 \\
$^{26}$Mg & 2.01E-02 & 3.18E-05 & 1.87E-07 & 6.25E-05 & 2.59E-05 & 1.41E-05 & 4.96E-05 & 2.97E-05 \\
$^{27}$Al & 1.48E-02 & 9.86E-04 & 1.13E-04 & 4.38E-04 & 2.47E-04 & 1.41E-04 & 4.45E-04 & 2.35E-04 \\
$^{28}$Si & 1.05E-01 & 1.54E-01 & 1.42E-01 & 2.72E-01 & 2.06E-01 & 1.58E-01 & 2.77E-01 & 1.98E-01 \\
$^{29}$Si & 8.99E-03 & 9.08E-04 & 5.79E-05 & 5.47E-04 & 3.40E-04 & 2.13E-04 & 5.52E-04 & 3.22E-04 \\
$^{30}$Si & 8.05E-03 & 1.69E-03 & 7.12E-05 & 1.03E-03 & 6.41E-04 & 3.88E-04 & 1.05E-03 & 6.14E-04 \\
$^{31}$P  & 1.21E-03 & 3.57E-04 & 9.12E-05 & 2.38E-04 & 1.60E-04 & 1.07E-04 & 2.40E-04 & 1.52E-04 \\
$^{32}$S  & 3.84E-02 & 8.46E-02 & 9.14E-02 & 1.60E-01 & 1.22E-01 & 9.37E-02 & 1.63E-01 & 1.17E-01 \\
$^{33}$S  & 1.78E-04 & 4.24E-04 & 6.07E-05 & 2.74E-04 & 1.92E-04 & 1.34E-04 & 2.71E-04 & 1.79E-04 \\
$^{34}$S  & 2.62E-03 & 1.98E-03 & 1.74E-05 & 2.76E-03 & 2.04E-03 & 1.46E-03 & 2.77E-03 & 1.90E-03 \\
$^{36}$S  & 1.78E-06 & 4.18E-07 & 3.41E-11 & 2.23E-07 & 1.31E-07 & 7.44E-08 & 2.22E-07 & 1.25E-07 \\
$^{35}$Cl & 1.01E-04 & 1.37E-04 & 1.06E-05 & 9.28E-05 & 7.07E-05 & 5.33E-05 & 9.03E-05 & 6.56E-05 \\
$^{37}$Cl & 1.88E-05 & 3.67E-05 & 5.56E-06 & 2.94E-05 & 2.26E-05 & 1.71E-05 & 2.86E-05 & 2.11E-05 \\
$^{36}$Ar & 6.62E-03 & 1.47E-02 & 1.91E-02 & 3.11E-02 & 2.41E-02 & 1.87E-02 & 3.18E-02 & 2.34E-02 \\
$^{38}$Ar & 1.37E-03 & 9.50E-04 & 6.60E-07 & 1.23E-03 & 9.90E-04 & 7.44E-04 & 1.20E-03 & 9.26E-04 \\
$^{40}$Ar & 2.27E-08 & 1.87E-08 & 3.42E-12 & 7.81E-09 & 5.19E-09 & 3.56E-09 & 7.56E-09 & 4.82E-09 \\
\hline
\end{tabular}
\end{center}
\end{table}

\newpage

\begin{table}
\centerline{Continued form TABLE 3.}
\vskip 0.2cm
\begin{center}
\begin{tabular}{lcccccccc}
\hline \hline
& \multicolumn{7}{c}{Synthesized mass (M$_\odot$)} \\
\cline{2-9} 
& \multicolumn{1}{c}{Type II} & \multicolumn{7}{c}{Type Ia} \\
Species & 10$-$50M$_\odot$ & W7 & W70 & WDD1 & WDD2 & WDD3 & CDD1 & CDD2 \\
\hline 
$^{39}$K  & 6.23E-05 & 7.23E-05 & 1.67E-06 & 6.76E-05 & 5.67E-05 & 4.52E-05 & 6.39E-05 & 5.34E-05 \\
$^{41}$K  & 5.07E-06 & 6.11E-06 & 4.83E-07 & 5.43E-06 & 4.52E-06 & 3.62E-06 & 5.20E-06 & 4.25E-06 \\
$^{40}$Ca & 5.77E-03 & 1.19E-02 & 1.81E-02 & 3.10E-02 & 2.43E-02 & 1.88E-02 & 3.18E-02 & 2.38E-02 \\
$^{42}$Ca & 4.23E-05 & 2.82E-05 & 1.06E-08 & 3.09E-05 & 2.55E-05 & 1.93E-05 & 2.97E-05 & 2.36E-05 \\
$^{43}$Ca & 1.08E-06 & 9.64E-08 & 6.17E-08 & 6.60E-08 & 2.22E-07 & 4.18E-07 & 5.15E-08 & 2.96E-07 \\
$^{44}$Ca & 5.53E-05 & 8.02E-06 & 1.38E-05 & 1.44E-05 & 2.95E-05 & 4.66E-05 & 1.37E-05 & 3.62E-05 \\
$^{46}$Ca & 1.43E-10 & 4.16E-09 & 1.01E-09 & 5.01E-09 & 4.73E-09 & 4.47E-09 & 8.79E-10 & 1.18E-09 \\
$^{48}$Ca & 5.33E-14 & 2.60E-09 & 2.47E-09 & 1.63E-06 & 1.64E-06 & 1.55E-06 & 3.54E-11 & 4.93E-10 \\
$^{45}$Sc & 2.29E-07 & 2.21E-07 & 3.85E-08 & 2.49E-07 & 2.09E-07 & 1.76E-07 & 2.47E-07 & 2.02E-07 \\
$^{46}$Ti & 7.48E-06 & 1.33E-05 & 3.49E-07 & 1.34E-05 & 1.12E-05 & 8.58E-06 & 1.27E-05 & 1.05E-05 \\
$^{47}$Ti & 2.11E-06 & 5.10E-07 & 4.08E-07 & 5.65E-07 & 1.56E-06 & 2.57E-06 & 4.93E-07 & 1.95E-06 \\
$^{48}$Ti & 1.16E-04 & 2.05E-04 & 3.13E-04 & 7.10E-04 & 6.11E-04 & 5.23E-04 & 7.32E-04 & 6.20E-04 \\
$^{49}$Ti & 5.98E-06 & 1.71E-05 & 2.94E-06 & 5.27E-05 & 4.39E-05 & 3.59E-05 & 5.22E-05 & 4.17E-05 \\
$^{50}$Ti & 3.81E-10 & 1.07E-04 & 1.04E-04 & 3.52E-04 & 3.51E-04 & 3.51E-04 & 2.08E-05 & 7.28E-05 \\
$^{50}$V  & 7.25E-10 & 1.55E-08 & 1.22E-08 & 9.74E-09 & 9.33E-09 & 9.07E-09 & 4.94E-09 & 1.20E-08 \\
$^{51}$V  & 1.00E-05 & 7.49E-05 & 4.27E-05 & 1.33E-04 & 1.16E-04 & 1.02E-04 & 1.11E-04 & 1.09E-04 \\
$^{50}$Cr & 4.64E-05 & 2.73E-04 & 6.65E-05 & 4.44E-04 & 3.53E-04 & 2.84E-04 & 4.49E-04 & 3.36E-04 \\
$^{52}$Cr & 1.15E-03 & 6.36E-03 & 7.73E-03 & 1.68E-02 & 1.37E-02 & 1.13E-02 & 1.65E-02 & 1.40E-02 \\
$^{53}$Cr & 1.19E-04 & 9.22E-04 & 5.66E-04 & 1.66E-03 & 1.38E-03 & 1.17E-03 & 1.59E-03 & 1.33E-03 \\
$^{54}$Cr & 2.33E-08 & 9.24E-04 & 9.04E-04 & 1.60E-03 & 1.60E-03 & 1.60E-03 & 2.31E-04 & 8.02E-04 \\
$^{55}$Mn & 3.86E-04 & 8.87E-03 & 6.66E-03 & 8.48E-03 & 7.05E-03 & 6.16E-03 & 8.10E-03 & 6.77E-03 \\
$^{54}$Fe & 3.62E-03 & 9.55E-02 & 7.30E-02 & 7.08E-02 & 5.91E-02 & 5.15E-02 & 7.20E-02 & 5.64E-02 \\
$^{56}$Fe & 8.44E-02 & 6.26E-01 & 6.80E-01 & 5.87E-01 & 7.13E-01 & 7.95E-01 & 5.65E-01 & 7.57E-01 \\
$^{57}$Fe & 2.72E-03 & 2.45E-02 & 1.92E-02 & 1.08E-02 & 1.67E-02 & 2.06E-02 & 1.01E-02 & 1.80E-02 \\
$^{58}$Fe & 7.22E-09 & 3.03E-03 & 2.96E-03 & 3.23E-03 & 3.23E-03 & 3.24E-03 & 8.63E-04 & 3.06E-03 \\
$^{59}$Co & 7.27E-05 & 1.04E-03 & 9.68E-04 & 3.95E-04 & 6.25E-04 & 7.75E-04 & 2.91E-04 & 6.35E-04 \\
$^{58}$Ni & 3.63E-03 & 1.10E-01 & 8.34E-02 & 3.14E-02 & 4.29E-02 & 4.97E-02 & 3.15E-02 & 4.47E-02 \\
$^{60}$Ni & 1.75E-03 & 1.24E-02 & 1.47E-02 & 5.08E-03 & 1.15E-02 & 1.67E-02 & 2.81E-03 & 1.21E-02 \\
$^{61}$Ni & 8.35E-05 & 2.35E-04 & 2.15E-04 & 7.00E-05 & 3.58E-04 & 5.92E-04 & 4.00E-05 & 4.27E-04 \\
\hline
\end{tabular}
\end{center}
\end{table}

\newpage

\begin{table}
\centerline{Continued form TABLE 3.}
\vskip 0.2cm
\begin{center}
\begin{tabular}{lcccccccc}
\hline \hline
& \multicolumn{7}{c}{Synthesized mass (M$_\odot$)} \\
\cline{2-9} 
& \multicolumn{1}{c}{Type II} & \multicolumn{7}{c}{Type Ia} \\
Species & 10$-$50M$_\odot$ & W7 & W70 & WDD1 & WDD2 & WDD3 & CDD1 & CDD2 \\
\hline 
$^{62}$Ni & 5.09E-04 & 3.07E-03 & 1.85E-03 & 1.37E-03 & 3.69E-03 & 5.46E-03 & 6.51E-04 & 4.60E-03 \\
$^{64}$Ni & 3.20E-14 & 1.70E-05 & 1.65E-05 & 2.32E-04 & 2.31E-04 & 2.32E-04 & 2.47E-06 & 9.29E-06 \\
$^{63}$Cu & 8.37E-07 & 2.32E-06 & 3.00E-06 & 2.77E-06 & 4.88E-06 & 5.92E-06 & 6.14E-07 & 3.97E-06 \\
$^{65}$Cu & 4.07E-07 & 6.84E-07 & 8.33E-07 & 7.08E-07 & 2.04E-06 & 3.38E-06 & 1.14E-07 & 2.05E-06 \\
$^{64}$Zn & 1.03E-05 & 1.06E-05 & 7.01E-05 & 3.71E-06 & 3.10E-05 & 5.76E-05 & 1.87E-06 & 3.96E-05 \\
$^{66}$Zn & 8.61E-06 & 1.76E-05 & 6.26E-06 & 2.16E-05 & 6.42E-05 & 1.04E-04 & 2.84E-06 & 6.11E-05 \\
$^{67}$Zn & 1.52E-08 & 1.58E-08 & 7.28E-09 & 6.35E-07 & 6.55E-07 & 6.27E-07 & 1.69E-09 & 4.01E-08 \\
$^{68}$Zn & 3.92E-09 & 1.74E-08 & 1.13E-08 & 7.44E-08 & 8.81E-08 & 9.42E-08 & 3.08E-09 & 3.03E-08 \\
\hline
\end{tabular}
\end{center}
\end{table}

\begin{table}
\centerline{Table 4: Major Radioactive Species in SN Ia Models}
\vskip 0.2cm
\begin{center}
\begin{tabular}{lccccccc}
\hline \hline
& \multicolumn{7}{c}{Synthesized mass (M$_\odot$)} \\
\cline{2-8} 
& \multicolumn{7}{c}{Type Ia} \\
Species & W7 & W70 & WDD1 & WDD2 & WDD3 & CDD1 & CDD2 \\
\hline 
$^{22}$Na & 1.73E-08 & 1.08E-08 & 2.66E-08 & 1.33E-08 & 3.04E-08 & 1.68E-08 & 9.
64E-09\\
$^{26}$Al & 4.93E-07 & 2.92E-08 & 4.61E-07 & 1.88E-07 & 1.64E-07 & 3.81E-07 & 2.32E-07\\
$^{36}$Cl & 2.58E-06 & 3.97E-10 & 1.25E-06 & 7.85E-07 & 4.89E-07 & 1.23E-06 & 7.35E-07\\
$^{39}$Ar & 1.20E-08 & 2.00E-13 & 6.04E-09 & 4.44E-09 & 3.29E-09 & 5.99E-09 & 4.12E-09\\
$^{40}$K  & 8.44E-08 & 5.46E-12 & 3.29E-08 & 2.35E-08 & 1.73E-08 & 3.13E-08 & 2.15E-08\\
$^{41}$Ca & 6.09E-06 & 4.83E-07 & 5.42E-06 & 4.52E-06 & 3.62E-06 & 5.20E-06 & 4.25E-06\\
$^{44}$Ti & 7.94E-06 & 1.38E-05 & 1.44E-05 & 2.95E-05 & 4.65E-05 & 1.36E-05 & 3.62E-05\\
$^{48}$V  & 4.95E-08 & 1.61E-08 & 6.52E-08 & 5.38E-08 & 4.11E-08 & 6.33E-08 & 5.03E-08\\
$^{49}$V  & 1.52E-07 & 3.23E-08 & 1.18E-07 & 1.03E-07 & 8.63E-08 & 9.99E-08 & 8.88E-08\\
$^{53}$Mn & 2.77E-04 & 2.48E-04 & 1.58E-04 & 1.54E-04 & 1.50E-04 & 8.47E-05 & 9.69E-05\\
$^{60}$Fe & 7.52E-07 & 7.19E-07 & 7.33E-05 & 7.33E-05 & 7.36E-05 & 6.23E-08 & 2.74E-07\\
$^{56}$Co & 1.44E-04 & 1.26E-04 & 6.04E-05 & 5.51E-05 & 5.20E-05 & 6.18E-05 & 5.24E-05\\
$^{57}$Co & 1.48E-03 & 1.42E-03 & 6.43E-04 & 6.42E-04 & 6.36E-04 & 4.50E-04 & 4.47E-04\\
$^{60}$Co & 4.22E-07 & 4.13E-07 & 3.76E-07 & 3.81E-07 & 3.78E-07 & 1.09E-07 & 4.23E-07\\
$^{56}$Ni & 5.86E-01 & 6.41E-01 & 5.64E-01 & 6.90E-01 & 7.73E-01 & 5.55E-01 & 7.37E-01\\
$^{57}$Ni & 2.27E-02 & 1.75E-02 & 9.95E-03 & 1.59E-02 & 1.98E-02 & 9.57E-03 & 1.74E-02\\
$^{59}$Ni & 6.71E-04 & 6.39E-04 & 2.54E-04 & 2.53E-04 & 2.51E-04 & 2.03E-04 & 1.90E-04\\
$^{63}$Ni & 8.00E-07 & 7.81E-07 & 1.69E-06 & 1.74E-06 & 1.73E-06 & 1.98E-07 & 8.31E-07\\
\hline
\end{tabular}
\end{center}
\end{table}
\newpage
\begin{table}
\centerline{Table 5: $N_{\rm Ia}/N_{\rm II}$ Ratios obtained for different Models of SN Ia}
\vskip 0.2cm
\label{tabIaII}
\begin{center}
\begin{tabular}{crrrrr}
\hline
SN II Model & W7 & W70 & DD1 & DD2 & DD3 \\
\hline
20~M$_\odot$ & 0.214 & 0.176 & 1.08 & 0.187 & 0.146 \\
10-50~M$_\odot$ & 0.295 & 0.243 & 1.48 & 0.257 & 0.201 \\
\hline
\end{tabular}
\end{center}
\end{table}

\end{document}